\documentclass[iop]{emulateapj}
\usepackage{graphicx}
\usepackage{color}
\usepackage[dvipsnames]{xcolor}
\usepackage{graphics}
\usepackage{amsmath}
\usepackage[colorlinks,citecolor=blue,urlcolor=blue,bookmarks=false,hypertexnames=true]{hyperref} 
\usepackage{natbib}
\usepackage{booktabs}
\usepackage[utf8]{inputenc}

\newcommand{\PerCubicGigaparsecPerYear}{Gpc$^{-3}$~yr$^{-1}$}
\newcommand{\Msol}{M$_\odot$}

\definecolor{ochre}{rgb}{0.8, 0.47, 0.13}

\usepackage[normalem]{ulem}

\shorttitle{Binary Black Hole Mass Distribution}
\shortauthors{S. Stevenson, M. Sampson, J. Powell et al.}

\begin{document}


\title{The impact of pair-instability mass loss on the binary black hole mass distribution}



\author{Simon Stevenson\altaffilmark{1,2}\email{spstevenson@swin.edu.au}}
\author{Matthew Sampson\altaffilmark{1,2,3}}
\author{Jade Powell\altaffilmark{1,2}}
\author{Alejandro Vigna-G\'{o}mez\altaffilmark{4,5}}
\author{Coenraad J. Neijssel\altaffilmark{4}}
\author{Dorottya Sz\'{e}csi\altaffilmark{4,6}}
\author{Ilya Mandel\altaffilmark{5,4,2}}

\affil{\altaffilmark{1} Center for Astrophysics and Supercomputing, Swinburne University of Technology, Hawthorn, VIC 3122, Australia.}
\affil{\altaffilmark{2} OzGrav: The ARC Center of Excellence for Gravitational Wave Discovery }
\affil{\altaffilmark{3} Queensland University of Technology, QLD , Brisbane 4000, Australia.}
\affil{\altaffilmark{4} School of Physics and Astronomy and Institute of Gravitational Wave Astronomy, University of Birmingham, Edgbaston, Birmingham B15 2TT, United Kingdom}
\affil{\altaffilmark{5} Monash Centre for Astrophysics, School of Physics and Astronomy,
Monash University, Clayton, Victoria 3168, Australia}
\affil{\altaffilmark{6} I. Physikalisches Institut, Universität zu Köln, Zülpicher-Strasse 77, D-50937 Cologne, Germany}

\begin{abstract}
A population of binary black hole mergers has now been observed in gravitational waves by Advanced LIGO and Virgo. The masses of these black holes appear to show evidence for a pile-up between $30$--$45$\,\Msol{} and a cut-off above $\sim 45$\,\Msol. One possible explanation for such a pile-up and subsequent cut-off are pulsational pair-instability supernovae (PPISNe) and pair-instability supernovae (PISNe) in massive stars. We investigate the plausibility of this explanation in the context of isolated massive binaries. We study a population of massive binaries using the rapid population synthesis software COMPAS, incorporating models for PPISNe and PISNe. Our models predict a maximum black hole mass of $40$\,\Msol{}. We expect $\sim 10$\% of all binary black hole mergers at redshift z = 0 will include at least one component that went through a PPISN (with mass $30$--$40$\,\Msol{}), constituting $\sim 20$--$50$\% of binary black hole mergers observed during the first two observing runs of Advanced LIGO and Virgo. Empirical models based on fitting the gravitational-wave mass measurements to a combination of a power law and a Gaussian find a fraction too large to be associated with PPISNe in our models. The rates of PPISNe and PISNe track the low metallicity star formation rate, increasing out to redshift $z = 2$. These predictions may be tested both with future gravitational-wave observations and with observations of superluminous supernovae. 
\end{abstract}

\section{Introduction}
\label{sec:intro}

Since beginning observing runs in 2015, Advanced LIGO \citep[aLIGO;][]{TheLIGOScientific:2014jea} and Advanced Virgo \citep[AdVirgo;][]{TheVirgo:2014hva} have confirmed the detection of 11 gravitational wave (GW) events to date \citep{Abbott:2016blz,Collaboration2018GWTC-1:Runsa}. The sources of 10 of the detections are binary black hole (BBH) mergers, and the source of the other event was the merger of a  binary neutron star (BNS) system  \citep{TheLIGOScientific:2017qsa,Collaboration2018GWTC-1:Runsa}. With the planned  advancements to current GW detectors around the world, the rate of GW detections is set to vastly increase over the coming years \citep{Aasi:2013wya}. GW observations will allow us to constrain astrophysical models \citep[e.g.][]{Stevenson:2015bqa,Zevin:2017evb,Wysocki:2017isg}.

The mass distribution of BBHs is commonly modelled as a power law \citep[e.g.][]{Fishbach:2017zga,2017PhRvD..95j3010K,Talbot:2018cva,Wysocki:2018mpo,Roulet:2018jbe,Collaboration2018BinaryVirgoa}, up to some maximum BH mass. With only 10 observed BBHs, the fine details of the BBH mass distribution remain uncertain at present \citep{Collaboration2018BinaryVirgoa}. However, even with a small number of detections, there is some evidence for certain features in the mass distribution.  

One of the most prominent features to emerge so far in the BBH mass distribution is a lack of BHs more massive than $\sim 45$\,\Msol{} \citep{Fishbach:2017zga,Talbot:2018cva,Wysocki:2018mpo,Roulet:2018jbe,Collaboration2018BinaryVirgoa}, despite ground-based GW detectors being sensitive to such systems. Currently, the most massive BH, observed in the event GW170729, had a mass of $50.6 ^{+16.6}_{-10.2}$\,\Msol{}   \citep{Collaboration2018GWTC-1:Runsa}. 


From the theory of stellar evolution, there is a prediction of an absence of BHs with masses $\sim 50$--$150$\,\Msol{}. Massive stars with helium core masses in the range $\sim 50$--$150$\,\Msol{} are believed to become unstable to electron-positron pair production \citep{FowlerHoyle:1964ApJS,1967PhRvL..18..379B,1968Ap&SS...2...96F}. This causes the radiation pressure support in the core to drop, causing the core to contract. As it contracts, the temperature increases, triggering explosive oxygen burning. This may reverse the contraction and completely unbind the star in a pair-instability supernova (PISN) explosion, leaving no remnant behind \citep[e.g.][]{2014A&A...565A..70K,2014A&A...566A.146K,2017MNRAS.464.2854K}. BH formation is expected again above a helium core mass of $\sim 150$\,M$_\odot$ \citep{Woosley2002TheStars}.

There is also tentative evidence for an excess of BHs with masses in the range $30$--$45$\,\Msol{} \citep{Collaboration2018BinaryVirgoa}. Theory suggests that stars with pre-supernova helium core masses in the range $\sim 30$--$50$\,\Msol{} also experience the pair-instability, but the release of energy is insufficient to completely disrupt the star. These stars may instead experience pair-instability induced pulsations multiple times \citep[e.g.][]{Woosley:2016hmi,Yoshida:2016MNRAS,Marchant:2018kun}. Each pulsation ejects material in a supernova-like event, leading to them being typically called pulsational pair-instability supernovae (PPISNe). 
After every pulse, the star's structure converges back to equilibrium. An iron core forms in equilibrium, and then collapses in a regular core-collapse supernova  \citep[CCSN;][]{Woosley:2007qp}. These pulsations remove the envelope of the star, rather than completely unbind the star. Enhanced mass loss during the pulsations may cause all stars in this mass range to form BH remnants with masses in a narrow range.

To date there have been no confirmed observations of PPISNe or PISNe, although some fraction of superluminous supernovae (SLSN-R) are potential candidates \citep{Woosley:2007qp,GalYam:2009Nature,Quimby:2011Nature,Cooke:2012Nature,2012Sci...337..927G,Lunnan:2016ApJ,Kozyreva:2018MNRAS,Gomez:2019yeb}. The supernova iPTF2014hls \citep{Arcavi2017Nature} may potentially be a PPISN \citep{Woosley:2018wuj}, possibly from a stellar merger product \citep{Vigna-Gomez:2019sky}. The supernova iPTF16eh \citep{Lunnan:2018NatAstr} also shows evidence for a massive shell of material ejected $\sim 30$ years prior to explosion, consistent with a PPISN.

There are several possible alternate explanations for the apparent excess of BHs in the mass range $30$--$45$~\Msol{}. Due to the relatively small number of observed BBH mergers, the statistical uncertainties in the inferred BBH mass distribution are still large, and a pure power law mass distribution is only mildly disfavoured \citep{Collaboration2018BinaryVirgoa}. It is also likely that the true mass distribution of BBHs in nature is not a perfect power law, and the preference for a model with more degrees of freedom than a power law may simply be highlighting this fact. 

Alternatively, finding clusters of GW observations in different parts of parameter space (e.g. a cluster of massive BHs) may indicate that multiple formation channels are contributing to the observed population \citep[e.g.][]{Powell:2019nmw}. 

BBHs are thought to form through several evolutionary channels \citep{Mandel:2018hfr}. All channels in which the BHs are of stellar origin will be subject to the effects of PISNe and PPISNe described above. These include classic isolated binary evolution \citep[e.g.][]{Belczynski:2016obo,Stevenson2017FormationEvolution,Kruckow:2018slo,Spera:2018wnw}, dynamical formation in dense stellar environments such as open clusters \citep{Rastello:2019,Chatterjee:2016thb}, globular or nuclear clusters \citep{1993Natur.364..421K,Sigurdsson:1993zrm,Rodriguez:2016avt,Fragione:2018vty}, active galactic nuclei \citep{Bartos:2016dgn,Stone:2016wzz}, formation in triple \citep{Antonini:2017ash,Rodriguez:2018jqu} or quadruple systems \citep{Fragione:2019hqt}, through wide binaries in the field \citep{Michaely:2019aet} or through quasi-chemically homogeneous evolution in close tidally locked binaries \citep{Mandel:2015qlu,Marchant:2016wow,Eldridge:2016ymr}. 

Although PISNe are still likely to operate in dense stellar environments such as globular clusters \citep{2018PhRvL.120o1101R}, imposing a maximum BH mass, it may be possible to form more massive BHs in these environments. For several of the observed BBH mergers (e.g. GW150914, GW170729), the merger product would be a BH in the PISN mass gap \citep[][]{Collaboration2018GWTC-1:Runsa}. It is possible that such so called second generation BHs could again participate in mergers, filling this gap \citep{2016ApJ...824L..12O,Samsing:2017xod,2017ApJ...840L..24F,Gerosa:2017kvu,2018PhRvL.120o1101R,Kimball:2019mfs}. Black holes formed from stellar merger products in star clusters may also populate the PISN mass gap \citep[e.g.][]{DiCarlo:2019pmf}.

More exotic scenarios for the origin of the currently observed BBHs include primordial BHs \citep{Bird:2016dcv}, or several of the events being a single  gravitationally lensed BBH merger \citep{Broadhurst:2019ijv}. The effects of PISNe are likely not relevant to these scenarios. 

In this paper we focus on BBHs formed through classical isolated binary evolution. We introduce prescriptions for modelling the effects of both PISNe and PPISNe on massive stars in our rapid binary evolution code COMPAS \citep{Stevenson2017FormationEvolution,2018MNRAS.481.4009V,Barrett:2017fcw,coen}. We calculate the expected distribution of BH masses, and show that we expect to observe a maximum BH mass of around $\sim 40$\,\Msol. Our models also show a mild excess of BHs in the $30$--$45$\,\Msol{} range due to PPISNe, with around $10$\% of BBHs merging at redshift $z = 0$ having at least one component which has undergone PPISN. Due to GW selection effects favouring more massive systems, this corresponds to around $20$--$50$\% of observed BBH mergers having at least one BH which has undergone a PPISN. This supports the hypothesis that the more massive black holes in the BBH mergers GW150914 and GW170729 may have formed in this way. We also calculate the volumetric rate of PISNe, PPISNe, and BBH mergers , showing that they increase from redshift $z = 0$ to $z = 2$, tracking the low metallicity cosmic star formation rate. 

In Section~\ref{sec:sim_params}, we give a brief description of the population synthesis code and set-up used in this study. In Section~\ref{sec:results} we calculate the volumetric rate of various astrophysical phenomena including PISNe, PPISNe and BBH mergers as a function of redshift. We also present the intrinsic mass distributions of BBHs merging at redshift $z=0$. In Section~\ref{sec:with_selection_effects} we present our predicted observed BBH mass distributions, incorporating GW selection effects. We discuss our results, compare with the literature and highlight limitations of this study in Section~\ref{sec:discussion}. We conclude in Section~\ref{sec:conclusion}.

\section{Binary population model}
\label{sec:sim_params}

We use the population synthesis software COMPAS \citep{Stevenson2017FormationEvolution,2018MNRAS.481.4009V,Barrett:2017fcw,coen} to study a population of massive isolated binaries. COMPAS models the evolution of binary systems, incorporating approximate prescriptions for stellar evolution \citep{1996MNRAS.281..257T,Hurley:2000pk}, stellar mass loss \citep{Vink:2001cg,Belczynski:2010ApJ}, mass transfer and common envelope evolution (see section~\ref{subsec:mass_transfer}),  supernovae (see sections~\ref{subsec:supernova} \& \ref{subsec:models}) and GW emission \citep{1964PhRv..136.1224P}. Our model neglects the effects of stellar rotation and tides. 

\subsection{Initial binary distributions}
\label{subsec:initial_distributions}

In COMPAS, binaries are Monte Carlo sampled from probability distributions of initial conditions based on astrophysical observations. We draw the mass of the initially more massive star from an initial mass function \citep[IMF;][]{Kroupa:2000iv} in the mass range $5$--$150$\,\Msol, whilst the secondary mass is determined from the mass ratio, drawn from a uniform distribution \citep{Sana:2012Sci}, with a lower limit of 0.1\,\Msol{}, since stars with masses less than $\sim 0.08$\,\Msol{} do not burn hydrogen \citep{1963ApJ...137.1121K,1963PThPh..30..460H}. We account for the low mass binaries and single stars we do not simulate in section~\ref{subsec:rates_methods}. We discuss the uncertainties inherent in our population synthesis approach in Section~\ref{sec:discussion}.

We draw the initial orbital separation of the binary from a uniform in log distribution with limits of 0.1 and 1000 AU \citep{Sana:2012Sci}. We assume all binaries are initially circular. 

We assume that these initial distributions are independent. \citet{Moe:2017ApJS} recently presented a distribution of correlated initial parameters and binary fractions based on a compilation of observations. Uncertainties in the initial distributions of massive binaries typically translate to uncertainties in predictions of BBH rates at the factor of 2 level and do not dominate, with the exception of the initial mass function \citep{deMink:2015yea,Klencki:2018zrz}. We explicitly demonstrate the uncertainty in our predictions due to uncertainty in the initial mass function in section~\ref{sec:results}.

We make the simplifying assumption, in the absence of compelling evidence to the contrary, that these initial distributions do not depend on metallicity or redshift. 

We use a grid of 50 metallicities distributed uniformly in log space between $Z_\mathrm{min} = 1 \times 10^{-4}$ and $Z_\mathrm{max} = 0.03$, set by the limits of the underlying stellar models \citep{Pols:1998MNRAS,Hurley:2000pk}. At each metallicity we model the evolution of $2 \times 10^5$ binary systems, for a total of $1 \times 10^{7}$ binaries.

In the following subsections we briefly summarise some of the most relevant details for the present study.

\subsection{Mass Transfer Stability and Common Envelope Evolution}
\label{subsec:mass_transfer}

Stars expand during their lifetimes. If the stellar radius exceeds the Roche-radius of the star it will transfer mass to its companion star. The stability of this mass transfer depends on the response of both the stellar radius and the Roche-radius of the star.
This stability is often defined in a radius-mass relationship (see for example \citet{1972AcA....22...73P}, \citet{1987HjellmingRadii})
\begin{equation}
    \zeta = \frac{d \ln{R}}{d \ln{m}} .
\end{equation} 
The mass transfer is dynamically stable if $\zeta_{*} > \zeta_{RL}$, where $\zeta_{*}$ is the adiabatic response of the star, 
and $\zeta_{RL}$ the response of the Roche-lobe.
A summary of our prescriptions for the stellar response at different evolutionary stages can be found in section 2.2.4 of \citet{2018MNRAS.481.4009V}. If the mass transfer is stable we
evolve the system in a similar manner to \citet{Hurley:2002rf} (see
also \citet{Stevenson2017FormationEvolution}).

When the mass transfer is dynamically unstable the system will start a common-envelope event \citep{1976IAUS...73...75P,2013A&ARv..21...59I}. In a common-envelope event, the envelope of the donor star engulfs the entire binary. In order to determine whether the binary survives the common-envelope event or results in a stellar merger, we use the ``$\alpha$-$\lambda$'' formalism \citep{1984ApJ...277..355W,1990ApJ...358..189D,2013A&ARv..21...59I}. This formalism compares the orbital energy to the energy needed to unbind the envelope. We assume that all the orbital energy goes into expelling the envelope ($\alpha_\mathrm{CE} = 1$). Our parametrization for the binding energy of the envelope (characterised by $\lambda$) is determined from the fitting formulae of \citet{2010ApJ...716..114X}, as in \texttt{StarTrack} \citep{Dominik:2012kk}. A requirement is that we know both the core and envelope mass of the donor star. Massive stars crossing the Hertzsprung gap may not have developed a clear core-envelope separation \citep{2013A&ARv..21...59I}. If this is not the case, common-envelope events from these donors could always result in a stellar merger. This uncertainty can have a large impact on predictions for the rate of merging binary black holes \citep[up to an order of magnitude c.f.][]{Dominik:2012kk}. We assume that all common-envelope events with Hertzsprung-gap donors result in stellar mergers \citep{Belczynski:2006zi}.

\subsection{Typical evolutionary channel for BBHs
}
\label{subsec:evol_channel}

There are multiple possible channels for forming a BBH from a massive isolated binary. Here we briefly describe the typical evolutionary stages involved. At present, COMPAS does not model the formation of BBHs in close, tidally locked binaries through chemically homogeneous evolution \citep[][]{Mandel:2015qlu,Marchant:2016wow}; work is underway to incorporate this into our model.

The dominant channels for forming BBHs in our models are detailed in \citet{Stevenson2017FormationEvolution} and \citet{coen}. We provide a brief summary here. A typical BBH is formed from an initially wide binary with an orbital period of a few hundred days, consisting of two massive O-type stars with zero-age main-sequence (ZAMS) masses greater than $\sim 20$\,\Msol{}.  As the initially more massive star (the primary) evolves off the main sequence first, its radius increases. As it does, the primary fills its Roche lobe \citep{Eggleton:1983ApJ} and begins to transfer mass to its companion, until all of its envelope is removed, leaving a helium star. The helium star may undergo PPISNe before finally collapsing to a BH. The initially less massive secondary star then evolves off the main sequence and begins transferring mass onto the BH. This can either be stable mass transfer \citep[e.g.][]{2017MNRAS.465.2092P,Heuvel:2017sfe,coen}, or can result in the binary undergoing common envelope evolution \citep[e.g.][]{Dominik:2012kk,Belczynski:2016obo,Stevenson2017FormationEvolution,Marchant:2018kun,coen}; both dramatically shrink the orbital period of the binary and removing the hydrogen envelope of the secondary star (see Section~\ref{subsec:mass_transfer} for more details). Finally, the secondary star may undergo PPISNe before collapsing to a BH, forming a BBH. The orbit of the BBH then slowly shrinks due to emission of GWs \citep[][]{1964PhRv..136.1224P} over a time period of up to the age of the universe.

\subsection{Supernovae}
\label{subsec:supernova}

Massive stars end their lives in supernovae, resulting in either the formation of a neutron star, a BH, or the complete destruction of the star. From radio observations of isolated Galactic pulsars, neutron stars are known to receive velocity kicks at birth of up to several hundred km s$^{-1}$ \citep{Hobbs:2005yx}. Our assumptions for neutron star natal kicks are described in Section 2.2.3 of \citet{2018MNRAS.481.4009V}. It is unclear both observationally and theoretically whether BHs also receive such kicks \citep{2012MNRAS.425.2799R,Janka:2013hfa,Mandel:2015eta,Repetto:2017gry}.  For canonical CCSNe we use the `delayed' supernova engine from \citet{Fryer:2012ApJ} to determine the remnant masses, with kicks drawn from a Maxwellian with a 1D dispersion of 265 km s$^{-1}$, as for neutron stars \citep{Hobbs:2005yx,2018MNRAS.481.4009V}. These kicks are then suppressed by the fraction of mass falling back on to the BH, as described in \citet{Fryer:2012ApJ}. In practice, this means that most BHs formed in our models receive no natal kick. 

\subsection{PPISN \& PISN Models}
\label{subsec:models}

In order to investigate the occurrence rates and properties of PPISNe and PISNe, we need to be able to relate the final properties of the remnant to those of its progenitor. This involves several assumptions and is very model dependent. In this section, we therefore describe several different models for determining the remnant mass $M_\mathrm{final}$ for BHs formed from progenitors earlier undergoing PPISN as a function of their pre-collapse helium core mass $M_\mathrm{He}$. These models are based on \citet{Belczynski:2016jno} (section~\ref{subsubsec:belczynski}), \citet{Woosley:2016hmi} (section~\ref{subsubsec:woosley}), \citet{Marchant:2018kun} (section~\ref{subsubsec:marchant}) and \citet{Woosley2019TheLoss} (section~\ref{subsubsec:woosley19}). Other studies of PPISNe in the literature include \citet{Leung:2019fgj}, which finds similar results to \citet{Marchant:2018kun} and  \citet{Woosley:2016hmi}, and \citet{2015A&A...573A..18M} and \citet{Yoshida:2016MNRAS}, which do not present a dense enough grid of models to be useful for the present study.

\begin{figure*}
\centering
\includegraphics[width=0.9\textwidth]{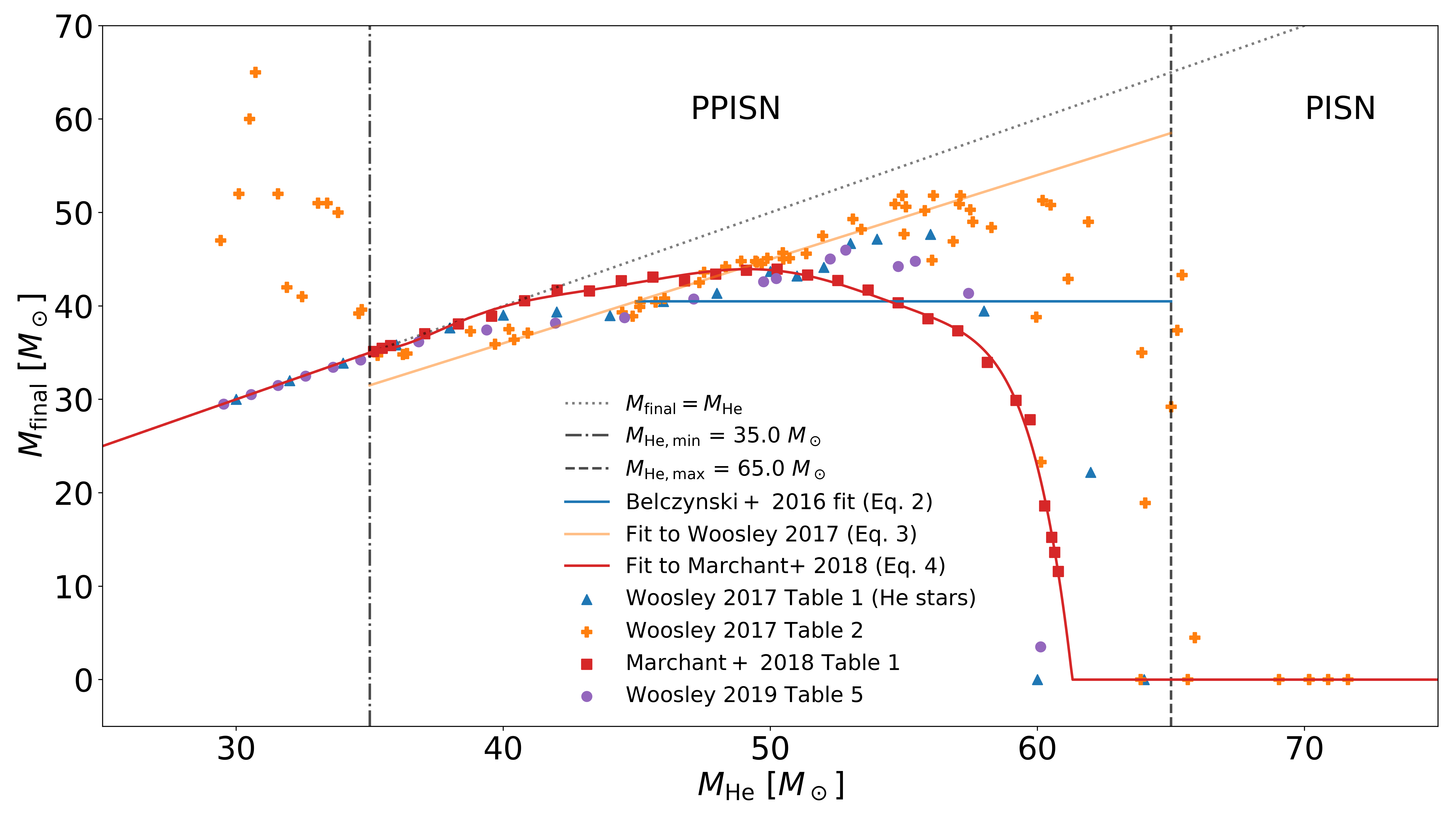}
\caption{Final remnant mass $M_\mathrm{final}$ as a function of pre-supernova helium (core) mass $M_\mathrm{He}$. Helium star models from Table 1 in \citet{Woosley:2016hmi} are plotted as blue triangles, whilst orange pluses are from the models in his Table 2. Models from Table 1 in \citet{Marchant:2018kun} are plotted as red squares. Models from Table 5 of \citet{Woosley2019TheLoss} are plotted in purple circles. The gray vertical lines denote the default boundaries between PPISN and PISN assumed by COMPAS \citep{Stevenson2017FormationEvolution}; these are adjustable by the user. The horizontal blue line shows the prescription used by \citet{Belczynski:2016jno} (Equation~\ref{eq:belczynski_2016_fit}), which is not a particularly accurate fit to the latest models. The solid orange line shows our linear fit to the \citet{Woosley:2016hmi} models given in Equation~\ref{eq:woosley_2017_fit}. The solid red line shows our polynomial fit to the models of \citet{Marchant:2018kun} given in Equation~\ref{eq:marchant_polynomial_fit}.}
\label{fig:mremvmcore}
\end{figure*}

For each model, we present a simple analytic fit to the results of the more detailed models. We apply this fit between the limiting values for the helium core mass given in Table~\ref{table:turnon_table}. We visualise both the original data and our fits in Figure~\ref{fig:mremvmcore}. In all models, we assume that stars with helium cores more massive than $M_\mathrm{He,max}$ and less massive than 135\,\Msol{} \citep{Woosley:2016hmi} undergo a PISN and leave no remnant. Cores more massive than 135\,\Msol{} are assumed to directly collapse to BHs \citep{Woosley2002TheStars,Woosley:2016hmi,Fryer:2012ApJ}. In this paper, we only study stars with ZAMS masses below 150\,$M_\odot$, and so we never form such massive cores. We leave an analysis of BBHs with component masses above the PISN mass gap to future work. See section~\ref{sec:discussion} for a discussion of issues with stellar models at such high masses. We do not explicitly model the multiple mass loss kicks \citep{1961BAN....15..265B,1961BAN....15..291B} from the multiple PPISNe mass loss episodes \citep{Woosley:2016hmi,Marchant:2018kun}; since the mass lost in most episodes is small, we do not expect this to significantly impact our results.

\begin{table}[]
\centering
\begin{tabular}{@{}lcc@{}}
\toprule
Model            & M$_\mathrm{He, min}$ & M$_\mathrm{He, max}$  \\ \midrule
Belczynski+ 2016 & 45        & 60        \\
Woosley 2017     & 35        & 65        \\
Marchant+ 2018   & 35        & 60        \\
Woosley 2019     & 30        & 60        \\
\bottomrule
\end{tabular}
\caption{Minimum and maximum helium core masses to undergo PPISN assumed in the various models}
\label{table:turnon_table}
\end{table}

\subsubsection{Belczynski et al. 2016 }
\label{subsubsec:belczynski}

\citet{Belczynski:2016jno} assume that all stars with a helium core more massive than 45\,\Msol{} lose all of their exterior mass through PPISN. They give the final BH mass as
\begin{equation}
    M_\mathrm{final} = 0.9 \times 45 ~\mathrm{M}_\odot .
\label{eq:belczynski_2016_fit}
\end{equation}
This prescription is not a particularly accurate approximation to the more detailed models discussed below.  
We include it here to demonstrate the effect of using this model on our predictions. This model has been adopted by other authors investigating the BH mass distribution \citep[e.g.][]{2018PhRvL.120o1101R}.

\subsubsection{Woosley 2017}
\label{subsubsec:woosley}

\citet{Woosley:2016hmi} simulates a grid of pure helium stars without mass loss with masses between 30 and 64\,\Msol{}. They also evolve a grid of hydrogen-rich stars with masses between 70 and 150\,\Msol{} and metallicity $Z = 0.1 \, Z_\odot$, including mass loss. We perform a linear regression to fit the remnant BH mass as a function of the pre-supernova helium core mass as
\begin{equation}
    M_\mathrm{final} = 0.9 \times M_\mathrm{He} .
\label{eq:woosley_2017_fit}
\end{equation}
This fit does not attempt to capture any possible turnover in the relation at around 60\,\Msol{}.

\subsubsection{Marchant et al. 2018}
\label{subsubsec:marchant}

\begin{table}[]
\centering
\begin{tabular}{@{}lr@{}}
\toprule
Coefficient & Value  \\ \midrule
$c_0$   &  7.39643451 $\times 10^{3}$  \\
$c_1$   & -1.13694590 $\times 10^{3}$  \\
$c_2$   &  7.45060098 $\times 10^{1}$  \\
$c_3$   & -2.69801221 $\times 10^{0}$ \\
$c_4$   & 5.83107626 $\times 10^{-2}$  \\
$c_5$   & -7.52206933 $\times 10^{-4}$  \\
$c_6$   & 5.36316755 $\times 10^{-6}$  \\
$c_7$   & -1.63057326 $\times 10^{-8}$  \\
\bottomrule
\end{tabular}
\caption{Coefficients for Equation~\ref{eq:marchant_polynomial_fit}}
\label{table:marchant_coefficients}
\end{table}

\citet{Marchant:2018kun} simulate a grid of single, non-rotating, pure helium stars with metallicity $Z = Z_\odot / 10$ with masses between 40 and 100\,\Msol{} using MESA \citep{Paxton:2011ApJS,Paxton:2017eie} including mass loss. These models demonstrate a turnover in the relation between pre-supernova helium core mass and final mass (see Figure~\ref{fig:mremvmcore}). We find that a 7$^\mathrm{th}$ order polynomial is a good fit to the data of \citet{Marchant:2018kun}, giving
\begin{equation}
    M_\mathrm{final} / M_\mathrm{He} = \sum_{\ell = 0}^{7} c_\ell \left( \frac{M_\mathrm{He}^\ell}{M_\odot} \right) .
\label{eq:marchant_polynomial_fit}
\end{equation}
The $c_\ell$ coefficients are given in Table~\ref{table:marchant_coefficients}. We bound the ratio between 0 and 1. We use this model as our reference model for Figures~\ref{fig:formation_efficiency_per_solar_mass_star_formation}, \ref{fig:rate_v_redshift}, \ref{fig:proportion_ppisn} \& \ref{fig:BBH_redshift_distributions}.


\subsubsection{Woosley 2019}
\label{subsubsec:woosley19}

\citet{Woosley2019TheLoss} also studies PPISNe in massive helium stars, including mass loss. The remnant masses calculated by \citet{Woosley2019TheLoss} agree reasonably well with the models of \citet{Marchant:2018kun}, although the former find a lower minium helium core mass which undergoes PPISN of $\sim 30$\,\Msol. Figure~\ref{fig:mremvmcore} shows that the models of \citet{Woosley2019TheLoss} are in broad agreement with those of \citet{Marchant:2018kun}. We therefore choose to use the same polynomial fitting formula given by Equation~\ref{eq:marchant_polynomial_fit}, with different values for $M_\mathrm{He, min}$ and $M_\mathrm{He, max}$ given in Table~\ref{table:turnon_table}.


\subsection{Caveats}
\label{subsubsec:cavet}

For our models based on both \citet{Marchant:2018kun} and \citet{Woosley2019TheLoss} we apply an additional factor of 0.9 to the masses to account for the difference between the baryonic and gravitational masses of the final BH, to be consistent with supernova engine adopted from \citet{Fryer:2012ApJ}. This avoids introducing an artificial mass gap at a mass $M_\mathrm{He, min}$ that we do not believe to be physical. However, the 10\% mass loss for BHs in \citet{Fryer:2012ApJ} is ad hoc and may be an overestimate\footnote{If the peak neutrino luminosity is $\sim 10^{53}$ erg s$^{-1}$, and the BH formation occurs on a timescale of $\sim 1$~s \citep{2016NCimR..39....1M,Chan:2017tdg} (substantially shorter than the Kelvin-Helmholtz timescale of the proto-neutron star), then the maximum energy loss can be a few $\times 10^{53}$ erg = 0.1~\Msol{} c$^2$.} (Bernhard M\"{u}ller, private communication, 2019). This will impact our estimates of the location of the maximum BH mass by up to $\sim 5$~\Msol{}.

Since the detailed PPISN models we use in Section~\ref{subsec:models} are only calculated for a single metallicity, we assume that the relation between pre-supernova helium core mass and BH mass is independent of metallicity. This can be updated in the future if significant metallicity effects are discovered in the detailed models.

In our models, we assume that the natal kick received by a BH formed from a progenitor which has undergone PPISNe is the same as one which did not.  In our \citet{Fryer:2012ApJ} natal kick prescription, this means that PPISN progenitors do not receive a natal kick in practice.  We do not model orbital changes from individual pulsations.

\section{Results}
\label{sec:results}

\begin{table*}[ht!]
\centering
\begin{tabular}{@{}ll@{}}
\toprule
Parameter            & Description \\ \midrule
$\mathcal{R}_\mathrm{BBH}$ & BBH Merger Rate in \PerCubicGigaparsecPerYear{} at $z = 0$ \\
$\mathcal{R}_\mathrm{BBH\,with\,PPISN}$     & PPISN Formed BBH Merger Rate in Gpc$^{-3}$yr$^{-1}$ at $z = 0$              \\
R$_\mathrm{BBH}$ & Predicted observed rate of BBH mergers during O1 and O2 per year \\
R$_\mathrm{BBH\,with\,PPISN}$ & Predicted observed rate of PPISN formed BBH mergers during O1 and O2 per year \\
$\Lambda_\mathrm{merge}$ & Proportion of BBHs merging at $z = 0$ formed via PPISNe \\
$\Lambda_\mathrm{obs}$ & Proportion of observed BBHs with at least one BH  formed from a progenitor which underwent PPISN \\
$M_\mathrm{max}$ & Max mass of PPISN-formed BH (\Msol) \\
$\beta$ & Max remnant mass formed not via PPISN (\Msol) \\
$Z_\mathrm{PPISN}$ & Highest metallicity PPISN formation occurred \\
$Z_\mathrm{PISN}$ & Highest metallicity PISN occurred \\
\bottomrule
\end{tabular}
\caption{Description of parameter names used in this paper}
\label{table:parameters}
\end{table*}

\begin{table*}[ht!]
\centering
\begin{tabular}{ l c c c c c c } 
\toprule
 Parameter & Woosley 2017 & Belczynski+ 2016 & Marchant+ 2018  & Woosley 2019 & IMF-2.1 & IMF-2.5 \\ 
\midrule
$\mathcal{R}_\mathrm{BBH}$ & 40 $\pm 4$ & 40 $\pm 4$ & 40 $\pm 4$ & 40 $\pm 4$ & 60 $\pm 6$ & 20 $\pm 2$ \\ 
$\mathcal{R}_\mathrm{BBH\,with\,PPISN}$ & 3 $\pm 1$ & 2 $\pm 1$ & 3 $\pm 1$ & 8 $\pm 2$ & 6 $\pm 3$ & 2 $\pm 1$ \\ 
R$_\mathrm{BBH}$  & 20 &  20 & 20 & 20 & 30 & 10  \\ 
R$_\mathrm{BBH\,with\,PPISN}$  & 7 & 3 & 5 & 8 & 10 & 3  \\ 
$\Lambda_\mathrm{merge}$ & 0.1 & 0.02 & 0.1 & 0.2 & 0.1  & 0.1 \\ 
$\Lambda_\mathrm{obs}$ & 0.4 & 0.2 & 0.4 & 0.5 & 0.4 & 0.3 \\ 
$M_\mathrm{max}$ & 58.4 & 40.5 & 39.5 & 39.5 & 58.4 & 58.4 \\ 
$\beta$ & 31.4 & 40.2 & 31.5 & 26.9 & 31.4 & 31.4 \\ 
$Z_\mathrm{PPISN}$ & 0.005 & 0.002 & 0.006 & 0.007 & 0.005 & 0.005 \\ 
$Z_\mathrm{PISN}$  & 0.001 & 0.001 & 0.002 &  0.002 & 0.001 & 0.001 \\ 
\bottomrule
\end{tabular}
\caption{Summary of results for each of our models described in section~\ref{subsec:models}. The uncertainties in the quoted rates are the 1$\sigma$ Monte-Carlo statistical uncertainties only. Descriptions of the parameters and their units are given in Table~\ref{table:parameters}.
\label{table:results}} 
\end{table*}

\begin{figure}
\centering
\includegraphics[width=\columnwidth]{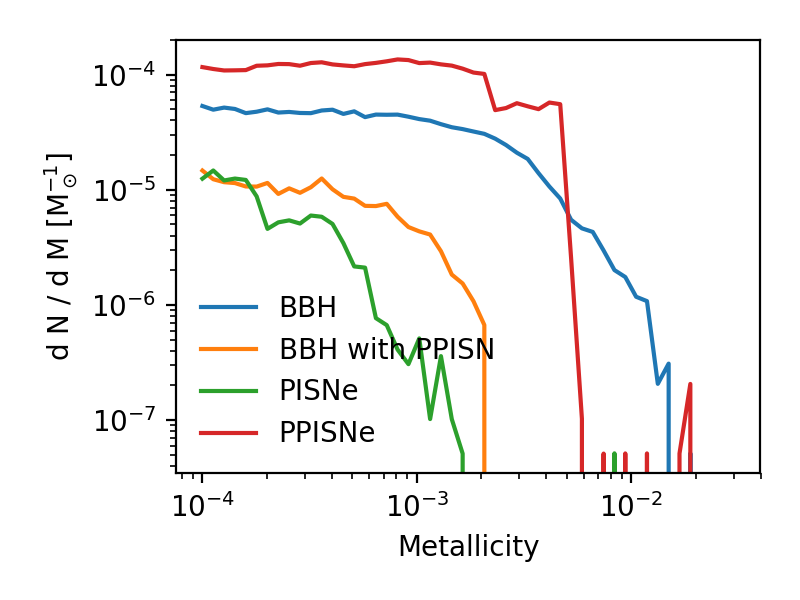}
\caption{Number of events formed per solar mass of star formation as a function of metallicity in our model. The blue line shows the total number of BBHs formed, whilst the orange line shows those BBHs where at least one of the BHs was formed from a progenitor which underwent PPISNe. The green line shows the number of PISNe, whilst the red line shows the same thing for PPISNe (see Section~\ref{sec:discussion} for caveats). The scatter between neighbouring metallicity bins gives an indication of sampling uncertainties. This figure uses our fit to the models of \citet{Marchant:2018kun} given in Equation~\ref{eq:marchant_polynomial_fit}; other models are qualitatively similar.}
\label{fig:formation_efficiency_per_solar_mass_star_formation}
\end{figure}

Massive stars lose a substantial fraction of their mass through line driven winds \citep{1975ApJ...195..157C} and eruptive mass loss \citep{1994PASP..106.1025H}. Line driven winds are thought to be quenched in low metallicity environments \citep{Vink:2001cg}. It is therefore expected that it is possible to form more massive helium cores, and thus more massive BHs, in lower metallicity environments \citep[e.g.][]{Belczynski:2010ApJ,Spera:2015MNRAS}. We show in Figure~\ref{fig:formation_efficiency_per_solar_mass_star_formation} that BBHs form at all metallicities $Z \lesssim 0.02$ in our models. Both PPISN and PISN occur down to our lowest metallicity of $Z = 1 \times 10^{-4}$, and we expect that they would continue to occur at lower metallicities \citep[e.g.][]{Fryer:2001ApJ,Yoon:2012A&A,Chatzopoulos:2012ApJ}. The highest metallicities at which PISNe and PPISNe occur in our models are around $Z \sim 0.002$ and $Z \sim 0.006$ respectively\footnote{We find a very small number of PISN and PPISN events at high metallicity $Z \gtrsim 0.01$ due to very particular mass transfer histories. Although we include these systems in computing rates, the maximum metallicities quoted in Table~\ref{table:results} exclude these rare systems} (see Figure~\ref{fig:formation_efficiency_per_solar_mass_star_formation} and Table~\ref{table:results} for details). These are in reasonable agreement with values found in previous studies for PISNe \citep{Belczynski:2016jno,Spera:2017fyx}.
However, \citet{Spera:2017fyx} find that PPISN occur in their models up to metallicities of $Z \sim 0.018$ (see their Fig.~3), much higher than the value we find. This discrepancy is likely due to different assumed stellar wind mass loss prescriptions, which are highly uncertain \citep{Renzo:2017}.

\subsection{Merger and Supernovae Rates - Methods}
\label{subsec:rates_methods}

We calculate the volumetric rates of 7 different astrophysical phenomena:
\begin{enumerate}
    \item the rate of stars experiencing PPISNe
    \item the rate of PISNe
    \item the total rate of CCSNe including those stars which went through PPISNe before collapsing
    \item formation rate of BBHs
    \item formation rate of BBHs in which at least one of the component BHs formed from a progenitor that underwent PPISN
    \item BBH merger rate
    \item rate of merging BBHs in which at least one of the component BHs formed from a progenitor that underwent PPISN.
\end{enumerate}
Each rate is calculated in a similar way. For example, the volumetric rate of PPISNe $\mathcal{R}_\mathrm{PPISN}$ at redshift $z$ is given by
\begin{equation}
\begin{aligned}
\mathcal{R}_\mathrm{PPISN}(z) = \frac{\mathrm{d}^2 N_\mathrm{PPISNe}}{\mathrm{d} t_\mathrm{s} \mathrm{d} V_\mathrm{c}} = \int \mathrm{d} Z \int \mathrm{d} \tau \\ \left[ \frac{\mathrm{d}^2 N_\mathrm{PPISNe}}{\mathrm{d} M_\mathrm{form} \mathrm{d} \tau} (Z) \frac{\mathrm{d}^3 M_\mathrm{form}}{\mathrm{d} t_s \mathrm{d} V_c \mathrm{d} Z} \left(Z, t_\mathrm{form} \left( \tau , z \right) \right) \right]
\label{eq:rateeq}
\end{aligned}
\end{equation}
%
where $Z$ is the metallicity and $t_\mathrm{form}$ is the age of the universe when a given source would need to have formed if there is delay $\tau$ between the time of formation and the event of interest.   

The first fraction in the square brackets in Equation~\ref{eq:rateeq} is the number of events (PPISNe) formed per unit star-forming mass per unit delay time at a metallicity Z. We show the number of events per unit star-forming mass as a function of metallicity in Figure~\ref{fig:formation_efficiency_per_solar_mass_star_formation}. In agreement with other authors \citep[e.g.][]{Giacobbo:2017qhh,Spera:2018wnw,coen} we find that the formation efficiency of BBHs is a strong function of metallicity, decreasing with increasing metallicity. At $Z = 1 \times 10^{-3}$, the formation efficiency of BBHs is $\sim 5 \times 10^{-5}$\,\Msol{}$^{-1}$. We also find that the formation efficiency of PISNe is a strong function of metallicity, decreasing from $\sim 10^{-5}$\,\Msol{}$^{-1}$ at $Z = 1 \times 10^{-4}$ to $\sim 10^{-6}$\,\Msol{}$^{-1}$ at $Z = 1 \times 10^{-3}$.

Since we did not include single stars\footnote{Our widest binaries have separations of up to $1000$~AU, and so do not interact} or low mass binaries (see Section~\ref{subsec:initial_distributions}), we must account for the mass we did not simulate. Using a Monte Carlo simulation with COMPAS over the full mass range 0.1--150\,\Msol{} and assuming a binary fraction of 70\% \citep{Sana:2012Sci} independent of primary mass, we find that we must multiply the total mass of our simulation by a factor of $\sim 5$ in order to account for the stars we did not simulate. Since we do not include single stars, our calculated rates are only for those events occurring in binaries. This means that the rates of supernovae in our model should be interpreted as lower limits. In addition, our adopted binary fraction is appropriate for the massive stars we are interested in, but we note that the binary fraction decreases with decreasing mass \citep[e.g.][]{Raghavan:2010ApJS,Moe:2017ApJS}. However, we do not expect this simplification to have a large impact on our results, as we focus on the outcome of the most massive stars' lives \citep[c.f.][]{Klencki:2018zrz}.

The second fraction in the square brackets in Equation~\ref{eq:rateeq} is the metallicity-specific star formation rate. Since we are interested in transient events with a strong metallicity dependence (as shown in Figure~\ref{fig:formation_efficiency_per_solar_mass_star_formation}), we need to know not only the total star formation rate, but the amount of star formation occurring at a given metallicity. Both the total star formation rate, and the fraction of star formation occurring at a given metallicity evolve with time (or equivalently redshift) throughout the history of the universe \citep{Madau:2014bja}. 

We assume an analytic, phenomenological model for the metallicity specific star formation rate introduced by \citet{coen}, calibrated to match GW observations. The star formation rate at redshift $z$ is
\begin{equation}
    \psi(z) = 0.01 \frac{(1+z)^{2.77}}{1 + ((1+z)/2.9)^{4.7}} \, \mathrm{M}_\odot \, \mathrm{Mpc}^{-3}\, \mathrm{yr}^{-1} ,
    \label{eq:star_formation_rate}
\end{equation}
where we take the star formation rate at redshift $z=0$ from \citet{Madau:2016jbv}. We assume that metallicities are log-normally distributed with a spread of 0.39 dex about a mean metallicity $\langle Z \rangle$ which scales with redshift $z$ as
\begin{equation}
    \langle Z \rangle =  \langle Z_{0} \rangle  \times 10^{-0.23 z}\, ,
    \label{eq:mean_metallicity_with_redshift}
\end{equation}
where $\langle Z_{0} \rangle = 0.035$ is the mean metallicity of the universe at redshift $z=0$.

We convert between lookback times and redshifts using the standard expressions \citep[e.g.][]{Hogg:1999ad}\footnote{We use a flat $\Lambda$-CDM cosmological model, using the WMAP9 cosmological parameters \citep{2013ApJS..208...19H} as implemented in \texttt{astropy}'s \texttt{cosmology} module. We convert between redshift and lookback time using \texttt{astropy}'s \texttt{lookback\_time}}.


Since the typical lifetime of massive stars is only a few Myr, we neglect any time between binary formation and supernovae ($\tau = 0$), and assume that these occur at the same redshift \citep[e.g.][]{Zapartas:2017zsb}. The PISN rate was calculated using the same method as above, with the substitution of $\mathrm{d} N_\mathrm{PISN}$ for $\mathrm{d} N_\mathrm{PPISNe}$ in Equation~\ref{eq:rateeq}. We do the same when calculating the BBH formation rate, substituting instead $\mathrm{d} N_\mathrm{BBH}$.


When calculating the BBH \emph{merger} rate, one must take more care, since the delay times $\tau$ between binary formation and GW driven merger can be long \citep[Gyrs; e.g.][]{Dominik:2012kk}. We therefore use $t_\mathrm{form} = t_\mathrm{merge}(z) - \tau$, where $t_\mathrm{merge}(z)$ is the age of the universe when the BHs merge \citep{Barrett:2017fcw}. Binaries with a merger time longer than the age of the universe at a given redshift are excluded from the analysis. We also calculate both the formation and merger rates of BBHs where at least one of the components was formed from a progenitor which underwent a PPISN. 

\subsection{Merger and Supernovae Rates - Results}
\label{subsec:rates_results}

\begin{figure}
\includegraphics[width=\columnwidth]{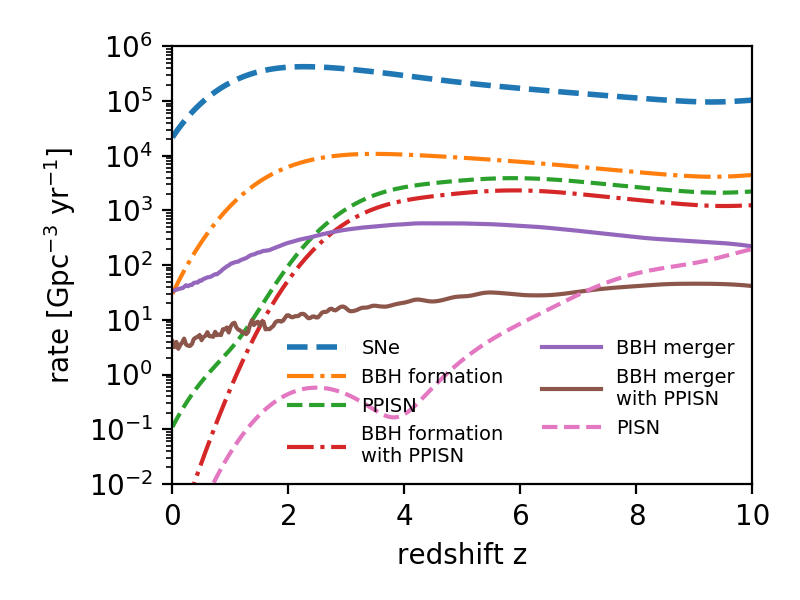}
\caption{Volumetric rates as a function of redshift. The dashed blue line shows the total core-collapse supernova rate. The dash-dot orange line shows the formation rate of all BBHs, including wide binaries that do not merge within the age of the universe. The dashed green line shows the rate of stars experiencing PPISNe. The dash-dot red line shows the formation rate of BBHs where at least one of the BHs was formed from a progenitor that went through a PPISN. The solid purple curve shows the BBH merger rate, whilst the solid brown curve shows the BBH merger rate where at least one BH was formed from a progenitor that went through PPISN. The dashed pink curve shows the volumetric rate of PISN. The rates for CCSNe, PISNe and PPISNe shown in this figure should all be treated as lower limits from our model, as discussed in  section~\ref{sec:discussion}. This figure uses our fit to the models of \citet{Marchant:2018kun} given in Equation~\ref{eq:marchant_polynomial_fit}; other models are qualitatively similar. 
}
\label{fig:rate_v_redshift}
\end{figure}

We show the rates of various astrophysical phenomena in Figure~\ref{fig:rate_v_redshift}. At redshift $z = 0$, the total CCSN rate is $\sim 2 \times 10^{4}$~\PerCubicGigaparsecPerYear.

Both PISNe and PPISNe are extremely rare events. At redshift $z=0$, we find the volumetric rate of PISNe to be less than $10^{-2}$~\PerCubicGigaparsecPerYear{} and the rate of PPISNe to be $\sim 10^{-1}$~\PerCubicGigaparsecPerYear{}. These events are rare at low redshift in our model because we assume that the mean metallicity of the local universe is super-solar (c.f. Equation~\ref{eq:mean_metallicity_with_redshift}).

At higher redshifts, where the mean metallicity of the universe is lower, the rates of these events increase. We find the rate of PPISNe at redshift $z = 4$ is $\sim 2 \times 10^{3}$~\PerCubicGigaparsecPerYear{} and the rate of PISNe at redshift $z = 4$ is $\sim 10^{-1}$~\PerCubicGigaparsecPerYear{}. As we discuss in section~\ref{sec:discussion}, our rate of PPISNe quoted here should be considered a lower limit.

We find that including PPISNe does not change the overall BBH merger rate appreciably compared to models not including PPISNe, in agreement with \citet{Belczynski:2016jno}. The BBH merger rate at redshift $z=0$ in our models is $\mathcal{R}_\mathrm{BBH} \sim 40$\,\PerCubicGigaparsecPerYear{} (see Table~\ref{table:results}), in agreement with empirical estimates from GW observations of $10$--$100$\,\PerCubicGigaparsecPerYear \citep{Collaboration2018BinaryVirgoa}. The total BBH merger rate is sensitive to the metallicity-specific star formation history of the Universe (see \citealt{coen} for a detailed discussion)
and additional assumptions about binary evolution (see Section~\ref{sec:discussion} for more details).  Therefore, we henceforth focus on the relative contribution of PPISN remnants as a fraction of all BBH mergers.

The rates of BBH mergers, PISNe and PPISNe all follow the low-metallicity star formation rate, initially increasing with redshift and peaking around redshift $z=2$ before decreasing to higher redshifts. This is due to a combination of both the overall star formation rate increasing, as well as the average metallicity decreasing between $z=0$ and $z=2$ \citep{Madau:2014bja}.

The BBH merger rate at a given redshift is typically an order of magnitude smaller than the BBH formation rate at the same redshift in our models, as a large fraction of BBHs have delay times longer than the age of the universe and never merge \citep[see also][]{Spera:2018wnw}. The BBH merger rate peaks at later times (lower redshifts) than the BBH formation rate due to the significant time delays between formation and merger.

\begin{figure}
\includegraphics[width=\columnwidth]{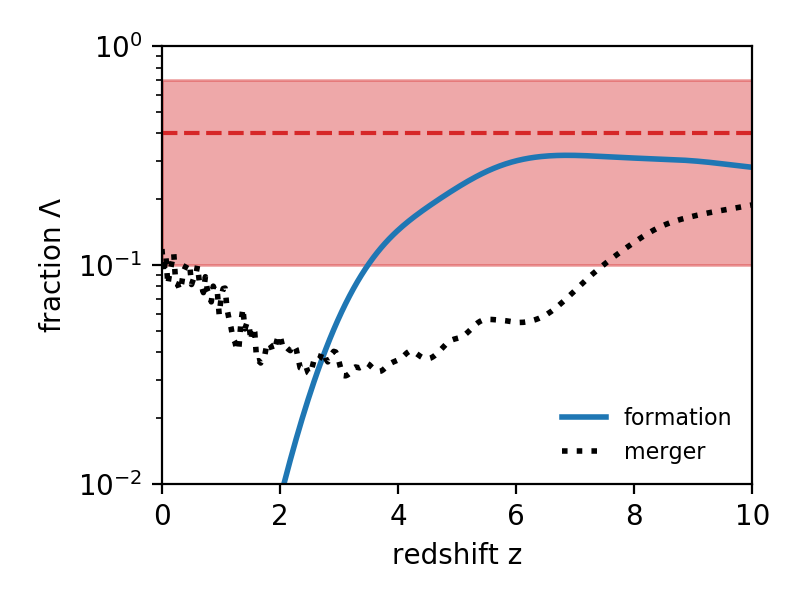}
\caption{The solid blue line shows the fraction of BBHs formed at a given redshift in which at least one of the BHs was formed from a progenitor that underwent PPISN, $\Lambda_\mathrm{form}$. The dotted black curve shows the proportion of BBHs merging at redshift $z$ in which at least one of the BHs was formed from a progenitor which underwent PPISN, $\Lambda_\mathrm{merge}$. The oscillations in this curve at low redshift are due to low number statistics. The dashed red line and shaded red area shows the median and 90\% credible region on the value inferred from gravitational-wave observations $\Lambda_\mathrm{LVC} = 0.4^{+0.3}_{-0.3}$ \citep{Collaboration2018BinaryVirgoa}. This figure uses our fit to the models of \citet{Marchant:2018kun} given in Equation~\ref{eq:marchant_polynomial_fit}; other models are qualitatively similar.}
\label{fig:proportion_ppisn}
\end{figure}

We define $\Lambda_\mathrm{merge}$ as the fraction of BBHs merging at a given redshift in which at least one BH was formed from a progenitor that went through a PPISN. We similarly define $\Lambda_\mathrm{form}$ to be the fraction of BBHs forming at a given redshift in which at least one BH was formed from a progenitor that went through a PPISN. We show the value for both $\Lambda_\mathrm{merge}$ and $\Lambda_\mathrm{form}$ as a function of redshift in Figure~\ref{fig:proportion_ppisn}. In our standard model based upon \citet{Marchant:2018kun}, around $\sim 10$\% of BBHs merging at redshift $z = 0$ have a component BH formed from a progenitor which underwent PPISN (see Table~\ref{table:results} for the value for other models). 

Our value for $\Lambda_\mathrm{merge}$ is towards the low end of the broad, empirically determined value $\Lambda_\mathrm{LVC} = 0.4^{+0.3}_{-0.3}$ \citep[][also shown in Figure~\ref{fig:proportion_ppisn}]{Collaboration2018BinaryVirgoa}. Our predicted rate of merging BBHs in which one of the BHs formed from a progenitor that went through PPISN is $\sim 3 $~\PerCubicGigaparsecPerYear, which is similarly lower than the rate of the high mass component inferred from GW observations of $10$--$40$~\PerCubicGigaparsecPerYear{} \citep{Collaboration2018BinaryVirgoa}\footnote{90\% credible interval obtained from publicly available posterior samples \url{https://dcc.ligo.org/LIGO-P1800324/public}}. 

This result suggests that standard binary evolution models may struggle to produce such a large fraction of BBHs through PPISNe, and another interpretation for the excess of massive black holes may be warranted. 

Our rates are sensitive to the assumed IMF \citep[e.g.][]{deMink:2015yea}, which is observationally uncertain \citep[e.g.][]{2018Sci...359...69S,FarrMandel:2018}.  To demonstrate this, we also include two models where we change the power law slope of the IMF for masses greater than 0.5\,\Msol{} from our default value of $-2.3$ \citep{Kroupa:2000iv} to $-2.1$ or $-2.5$, using our PPISN model based on \citet{Marchant:2018kun} (Equation~\ref{eq:marchant_polynomial_fit}). We find that varying the IMF in this range leads to a factor of 2 uncertainty in the total BBH merger rate at redshift $z=0$ \citep[see also][]{deMink:2015yea}. The fraction of BBH mergers at redshift $z=0$ in which at least one component underwent a PPISN is relatively robust to uncertainties in the IMF (see Table~\ref{table:results}).



\subsection{Maximum BH mass}
\label{subsec:upper_mass_limit}

The maximum BH mass $m_\mathrm{max}$ for each of our models is given in Table~\ref{table:results}. Our models based on \citet{Belczynski:2016jno}, \citet{Marchant:2018kun} and \citet{Woosley2019TheLoss} have an upper mass limit of $\sim 40$\,\Msol{} which is consistent with the maximum BH mass inferred from GW observations \citep{Fishbach:2017zga,Talbot:2018cva,Wysocki:2018mpo,Roulet:2018jbe,Collaboration2018BinaryVirgoa}.  


Our fit to the models of \citet{Woosley:2016hmi} results in a maximum BH mass of $\sim 53$\,\Msol{}. This higher limit can be explained by the fact that our fit assumes that the remnant mass increases linearly with increasing helium core mass, resulting in larger remnant masses $>50$\,\Msol{} close to maximum helium core mass $M_\mathrm{He, max}$. This can be seen in Figure~\ref{fig:mremvmcore}. 

Excluding the \citet{Belczynski:2016jno} model, the most massive BHs to be formed from progenitors without undergoing a PPISN ($\beta$ in Table~\ref{table:results}) are in the range of $\sim 27-32$\,\Msol{} for all our models, with exact numbers shown in Table~\ref{table:results}. In these models, BHs with masses in the range $35$--$45$\,\Msol{} are exclusively formed from progenitors that previously underwent PPISN events.

\subsection{BBH mass distribution}
\label{subsec:mass_spec}

\begin{figure*}
\includegraphics[width=0.5\textwidth]{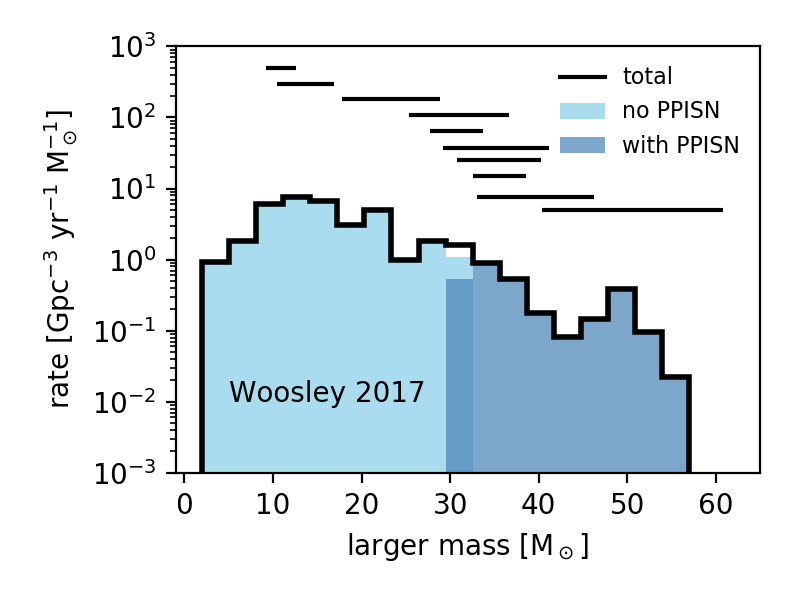}
\includegraphics[width=0.5\textwidth]{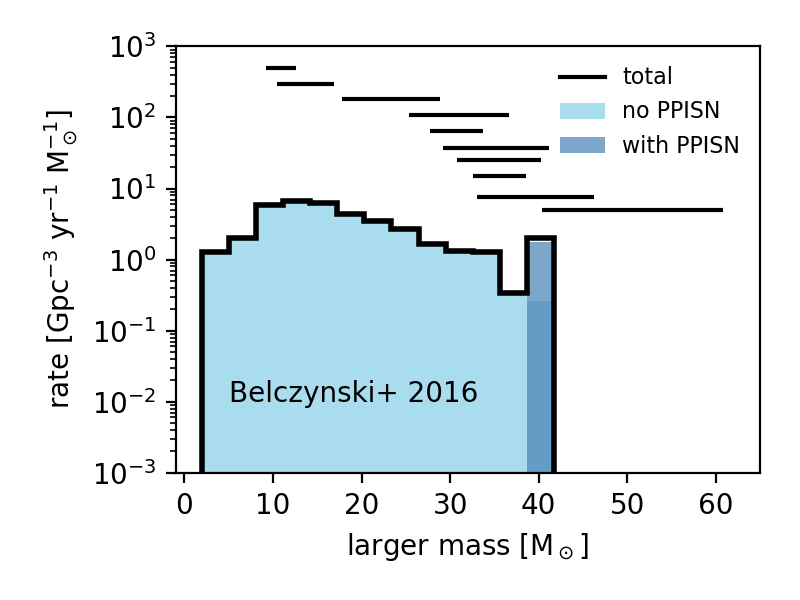}
\includegraphics[width=0.5\textwidth]{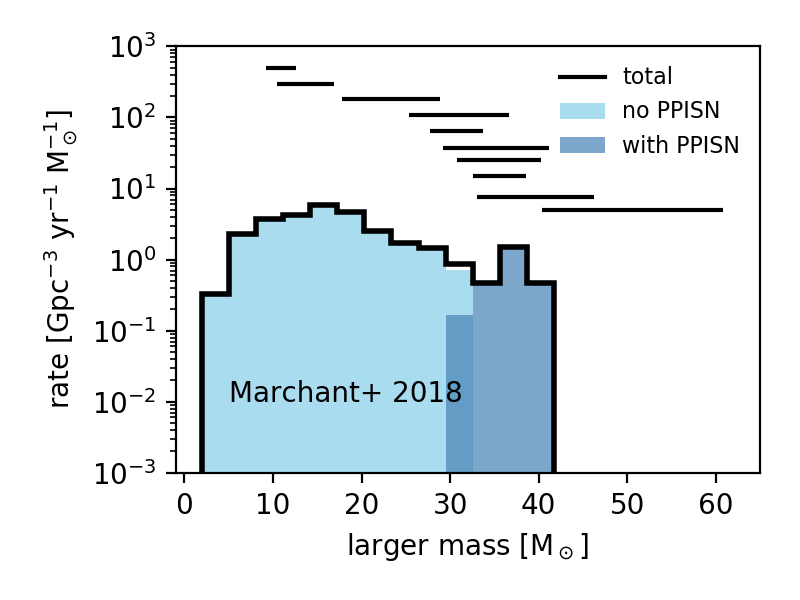}
\includegraphics[width=0.5\textwidth]{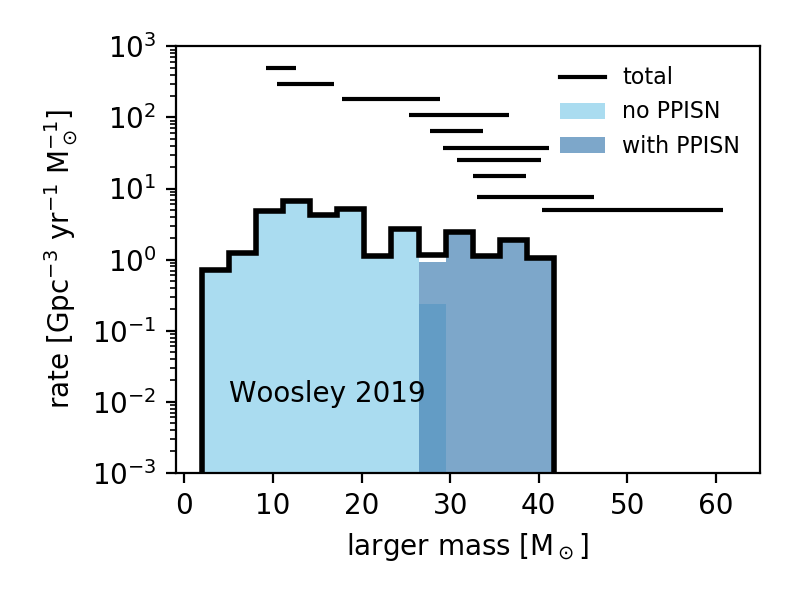}
\caption{Mass distribution of the more massive black hole in BBH mergers at redshift $z= 0$ predicted by our 4 models (see section~\ref{subsec:models}). The top left panel uses our linear fit to \citet{Woosley:2016hmi}, the top right uses the fit of \citet{Belczynski:2016jno}, the bottom left uses our fit to the models of \citet{Marchant:2018kun} and the bottom right uses the same fit applied to the models of \citet{Woosley2019TheLoss}. In each panel, the lighter blue shows the distribution of the mass of the more massive black hole in binaries with neither black hole formed through PPISN. The darker blue histogram shows the same quantity for binaries where at least one of the black holes formed from a progenitor which underwent a PPISN. The solid black histogram shows the total of these two. The horizontal black error bars show the gravitational-wave BBH observations from O1 and O2 \citep{Collaboration2018GWTC-1:Runsa,Collaboration2018BinaryVirgoa}; their vertical positions are arbitrary. GW observations are shown to provide context only and no direct comparison should be made, as this plot does not include GW selection effects (see Figure~\ref{fig:BBH_larger_mass_observed} for a plot including selection effects).  
}
\label{fig:BBH_larger_mass_spectrum}
\end{figure*}

\begin{figure*}
\includegraphics[width=0.5\textwidth]{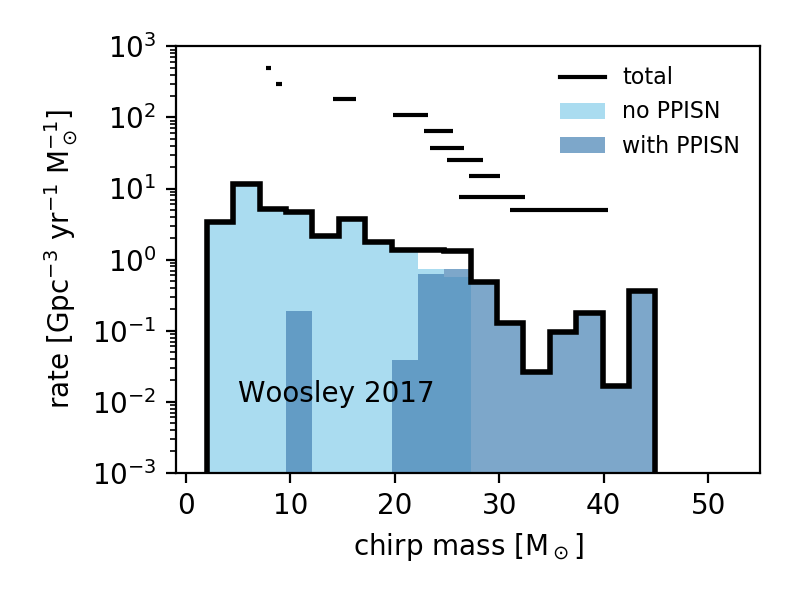}
\includegraphics[width=0.5\textwidth]{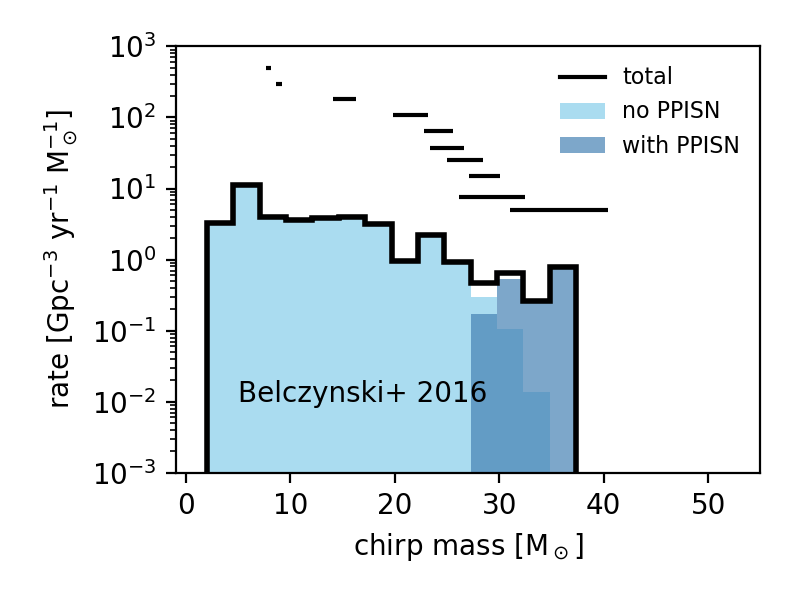}
\includegraphics[width=0.5\textwidth]{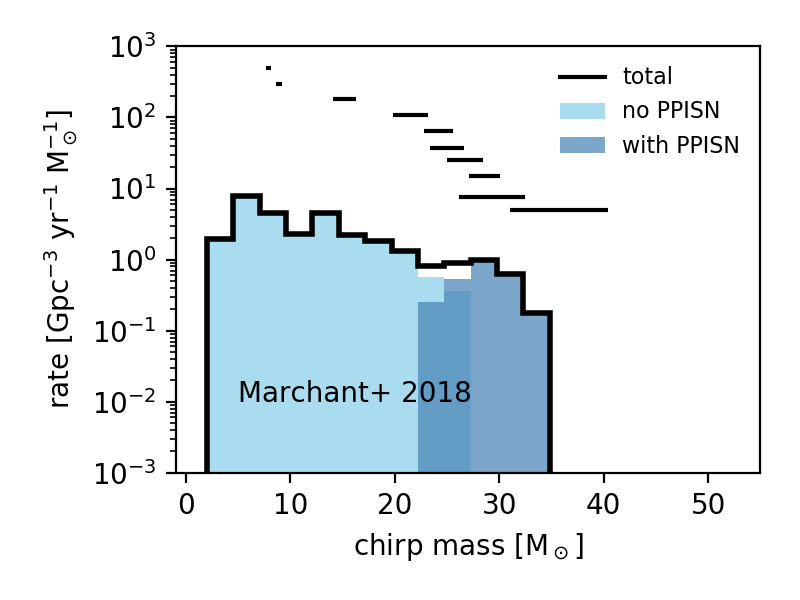}
\includegraphics[width=0.5\textwidth]{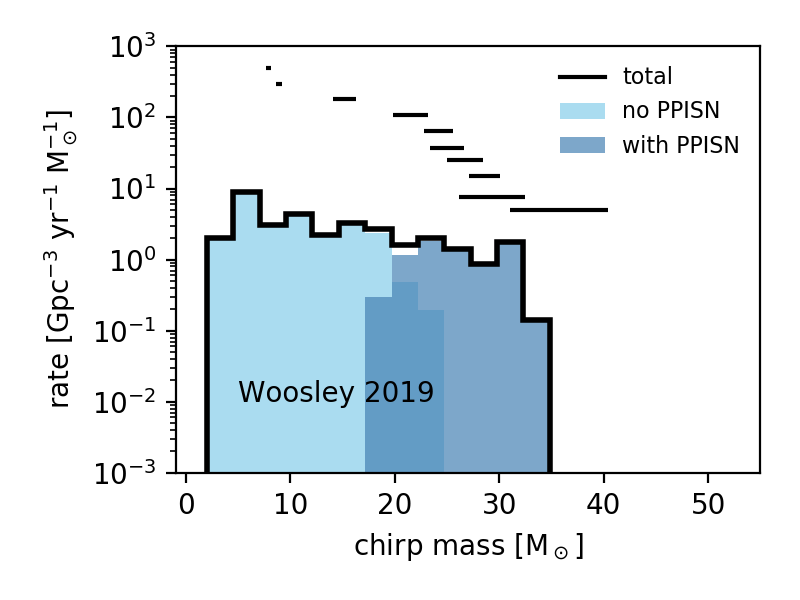}
\caption{Distribution of chirp masses in BBH mergers at redshift $z= 0$ predicted by our 4 models (see section~\ref{subsec:models}). The top left panel uses our linear fit to \citet{Woosley:2016hmi}, the top right uses the fit of \citet{Belczynski:2016jno}, the bottom left uses our fit to the models of \citet{Marchant:2018kun} and the bottom right uses the same fit applied to the models of \citet{Woosley2019TheLoss}. In each panel, the lighter blue shows the distribution of the chirp mass of binary black holes with neither black hole formed through PPISN. The darker blue histogram shows the same quantity for binaries where at least one of the black holes formed from a progenitor which underwent a PPISN. The solid black histogram shows the total of these two. The horizontal black error bars show the gravitational-wave binary black hole observations from O1 and O2 \citep{Collaboration2018GWTC-1:Runsa,Collaboration2018BinaryVirgoa}; their vertical positions are arbitrary. GW observations are shown to provide context only and no direct comparison should be made, as this plot does not include GW selection effects (see Figure~\ref{fig:BBH_chirp_mass_observed} for a plot including selection effects).
}
\label{fig:BBH_chirp_mass_spectrum}
\end{figure*}

We show the intrinsic BBH mass distribution in Figures~\ref{fig:BBH_larger_mass_spectrum} and \ref{fig:BBH_chirp_mass_spectrum}. Figure \ref{fig:BBH_larger_mass_spectrum} shows the intrinsic distribution of the mass of the more massive BH in a merging BBH at redshift $z=0$ for 4 of our models. The sharp spike at $\sim 41$\,\Msol{} in the \citet{Belczynski:2016jno} model is consistent with the features of the \citet{Belczynski:2016jno} fit shown in Figure~\ref{fig:mremvmcore}, where a flattening of the predicted remnant mass occurs at $\sim 41$\,\Msol{}.


The distribution is not consistent with a power law in mass across the entire mass range. Our model populations based on \citet{Woosley:2016hmi}, \citet{Marchant:2018kun} and \citet{Woosley2019TheLoss} produce BH mass distributions (for the more massive BH) which rise from 5 to $\sim 15$\,\Msol{}, and do not resemble a power law distribution above $\sim 15$\,\Msol{}. Above $\sim 34$\,\Msol{} there is an increase in the amount of BBHs in the $35-40$\,\Msol{} region peaking at a final remnant mass of $\sim 39$\,\Msol{}.

A well measured combination of the component masses in a BBH is the chirp mass
\begin{equation}
    \mathcal{M} = (m_1 m_2)^{3/5} / (m_1+m_2)^{1/5} .
    \label{eq:chirp_mass}
\end{equation}
We show the intrinsic chirp mass distributions predicted by our models of BBHs merging at redshift $z=0$ in Figure~\ref{fig:BBH_chirp_mass_spectrum}. Our models show that even BBHs with chirp masses as low as $\sim 10$\,\Msol{} may have one component BH formed from a progenitor which underwent a PPISN (in a binary with a lower mass BH). We discuss the predicted observed mass distributions in Section~\ref{sec:with_selection_effects}. The models presented in this section can be used to construct phenomenological models of the BBH mass distribution and provide physically motivated priors for their parameters.  

\begin{figure}
\includegraphics[width=\columnwidth]{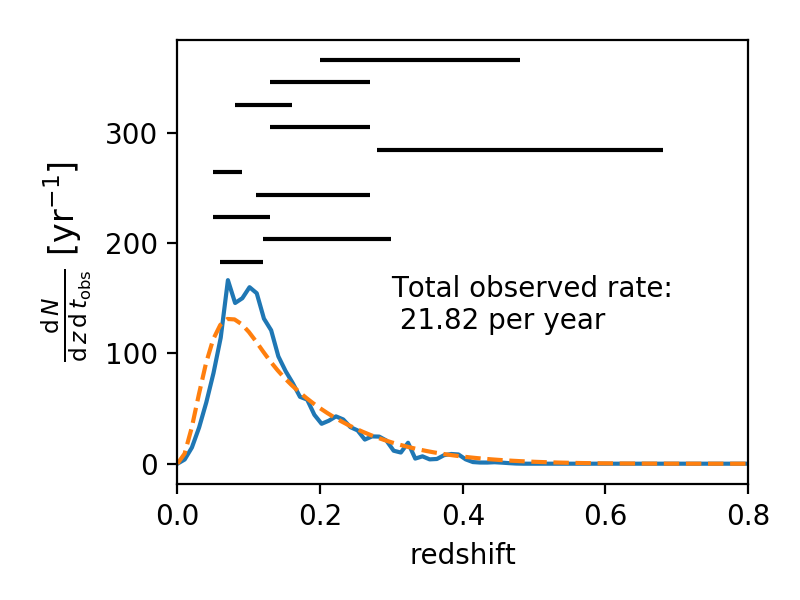}
\caption{Distribution of redshifts of BBH mergers observed in O1/O2 predicted by our model (blue solid). The maximum observed redshift predicted by our model is $\sim 0.6$. The horizontal black error bars show the gravitational-wave binary black hole observations from O1 and O2 \citep{Collaboration2018GWTC-1:Runsa,Collaboration2018BinaryVirgoa}; their vertical positions are arbitrary. For comparison, the dashed orange curve shows the distribution of detected redshifts for the analytic power-law model from Figure~10 in \citet{Collaboration2018BinaryVirgoa}, rescaled to the same number of observations per year. This figure uses our fit to the models of \citet{Marchant:2018kun} given in Equation~\ref{eq:marchant_polynomial_fit}; other models are qualitatively similar. 
}
\label{fig:BBH_redshift_distributions}
\end{figure}

\begin{figure*}
\includegraphics[width=0.5\textwidth]{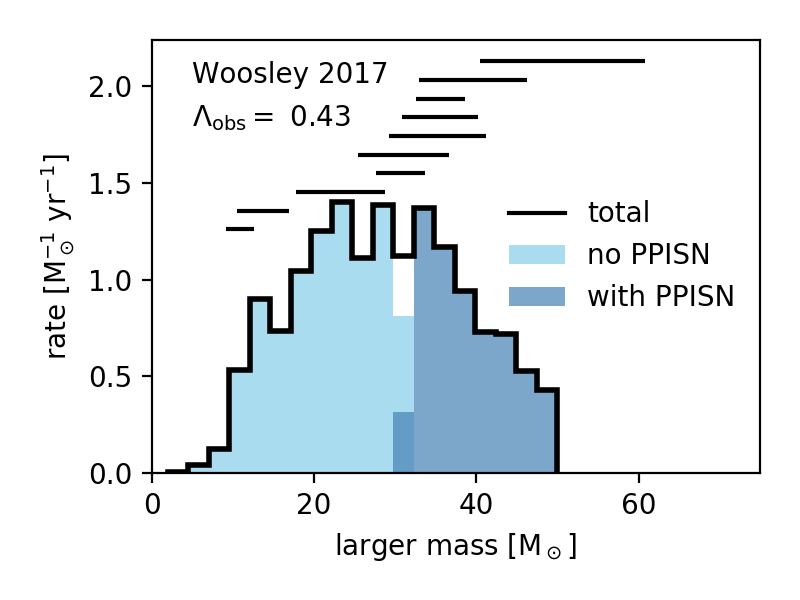}
\includegraphics[width=0.5\textwidth]{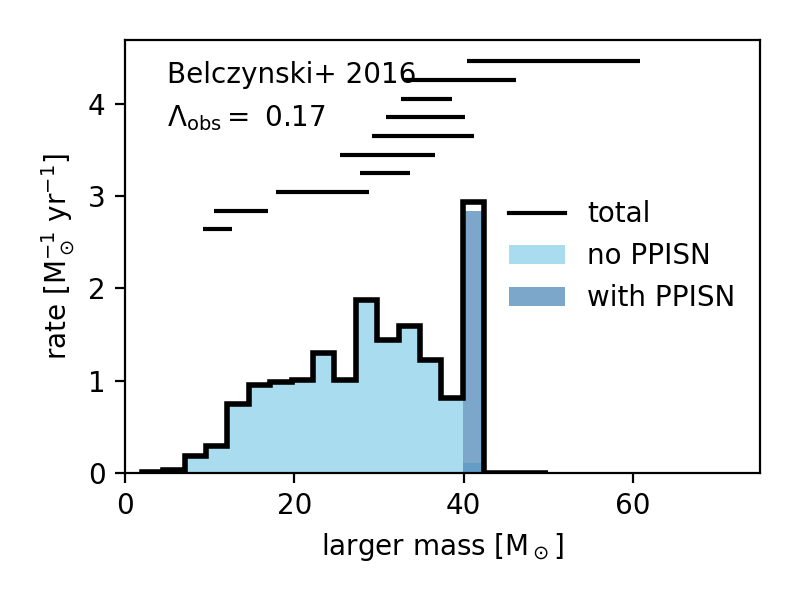}
\includegraphics[width=0.5\textwidth]{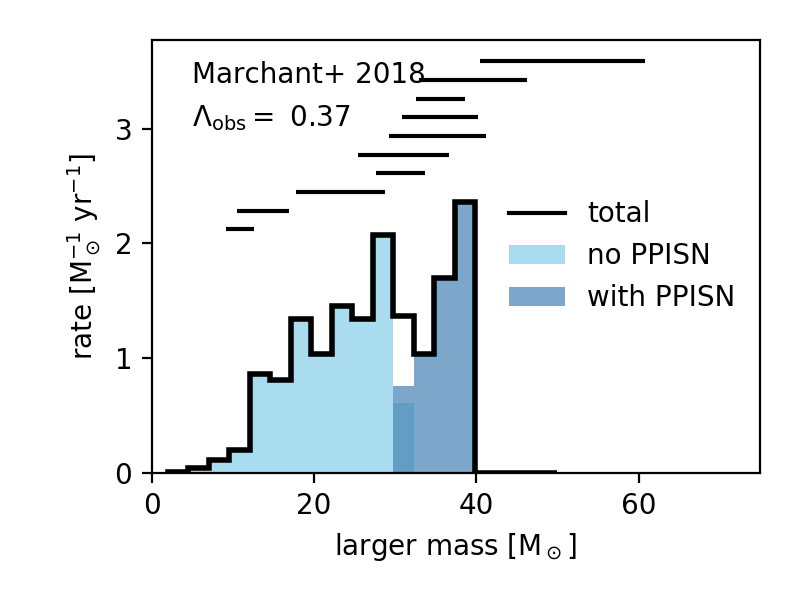}
\includegraphics[width=0.5\textwidth]{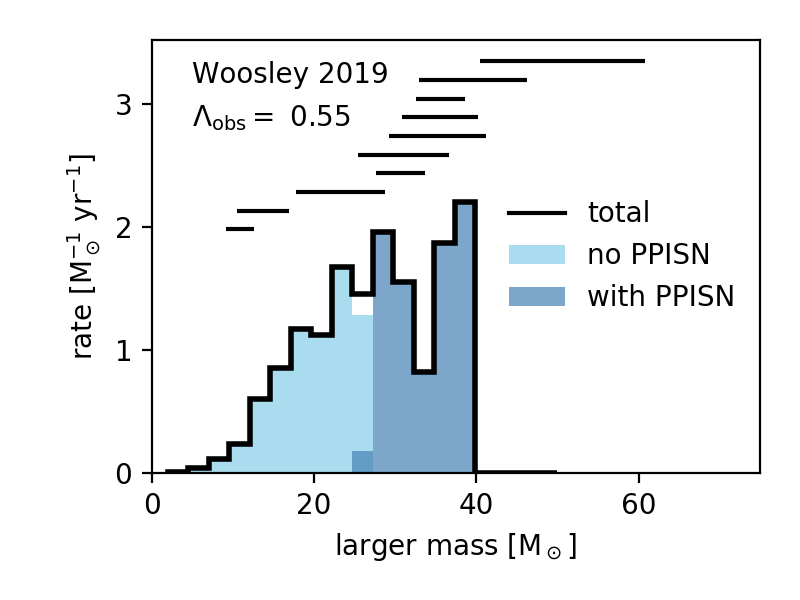}
\caption{Mass distributions for O1/O2 of the more massive black hole in observed BBH mergers predicted by our 4 models for O1/O2 (see section~\ref{subsec:models}). The top left panel uses our linear fit to \citet{Woosley:2016hmi}, the top right uses the fit of \citet{Belczynski:2016jno}, the bottom left uses our fit to the models of \citet{Marchant:2018kun} and the bottom right uses the same fit applied to the models of \citet{Woosley2019TheLoss}. In each panel, the lighter blue shows the distribution of the chirp mass of binary black holes with neither black hole formed through PPISN. The darker blue histogram shows the same quantity for binaries where at least one of the black holes underwent a PPISN; the fraction of observed BBHs where at least one of the black holes underwent a PPISN is shown on the plots labelled $\Lambda_\mathrm{obs}$.  The solid black histogram shows the total distribution.  The horizontal black error bars show the gravitational-wave binary black hole observations from O1 and O2 \citep{Collaboration2018GWTC-1:Runsa,Collaboration2018BinaryVirgoa}; their vertical positions are arbitrary. 
}
\label{fig:BBH_larger_mass_observed}
\end{figure*} 

\begin{figure*}
\includegraphics[width=0.5\textwidth]{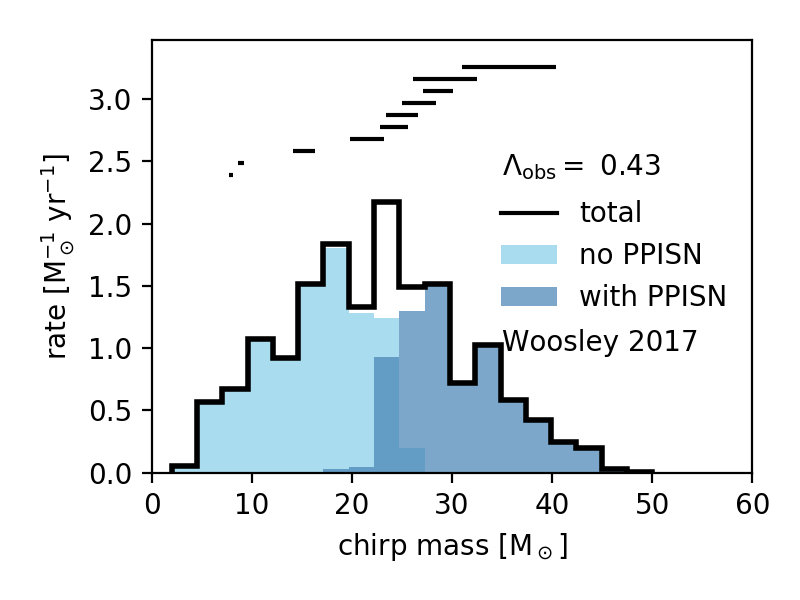}
\includegraphics[width=0.5\textwidth]{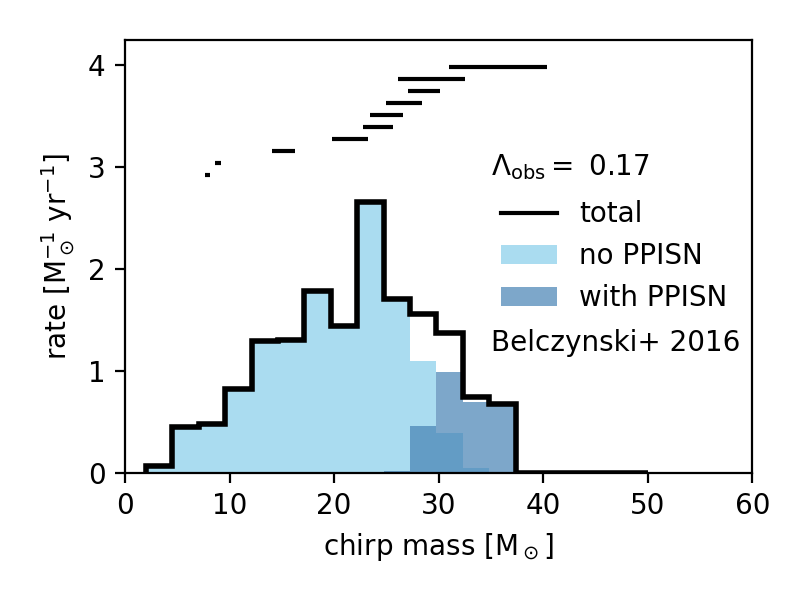}
\includegraphics[width=0.5\textwidth]{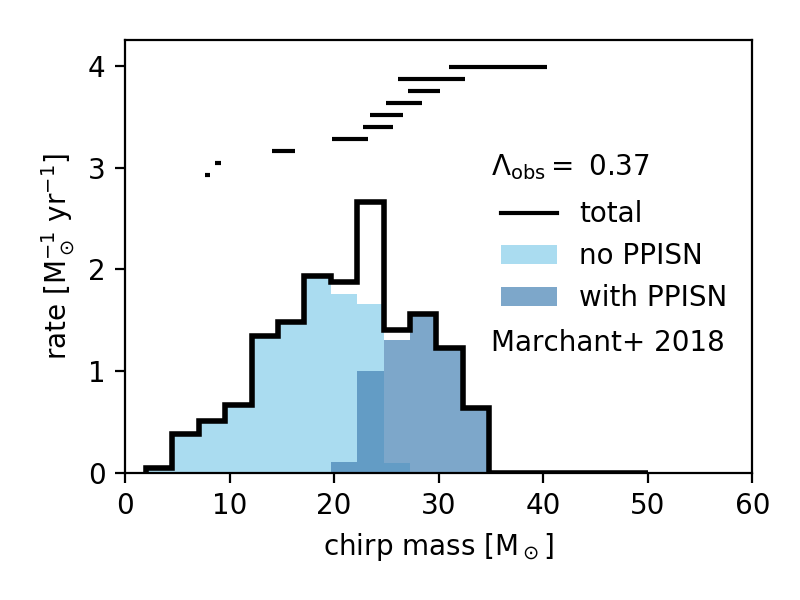}
\includegraphics[width=0.5\textwidth]{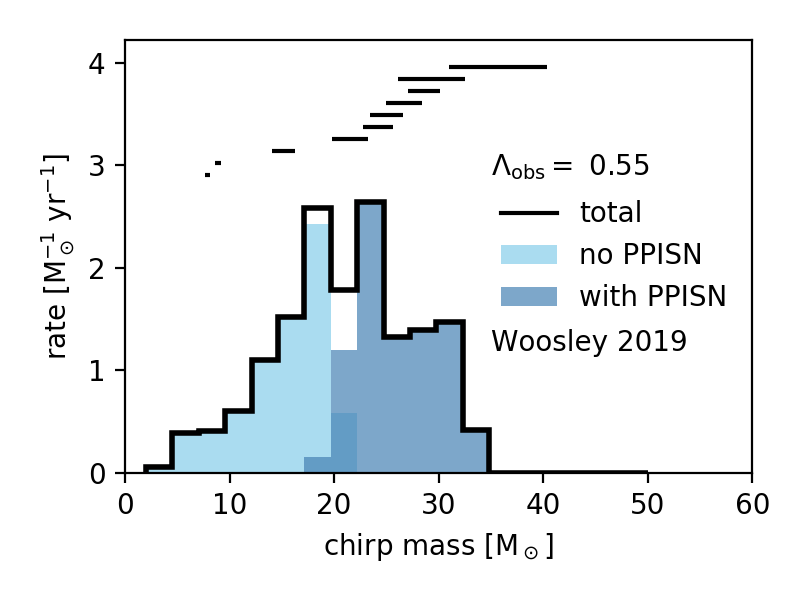}
\caption{Distribution of chirp masses of BBH mergers observed in O1/O2 predicted by our 4 models (see section~\ref{subsec:models}). The top left panel uses our linear fit to \citet{Woosley:2016hmi}, the top right uses the fit of \citet{Belczynski:2016jno}, the bottom left uses our fit to the models of \citet{Marchant:2018kun} and the bottom right uses the same fit applied to the models of \citet{Woosley2019TheLoss}. In each panel, the lighter blue shows the distribution of the chirp mass of binary black holes with neither black hole formed through PPISN. The darker blue histogram shows the same quantity for binaries where at least one of the black holes underwent a PPISN; the fraction of observed BBHs where at least one of the black holes underwent a PPISN is shown on the plots labelled $\Lambda_\mathrm{obs}$.  The solid black histogram shows the total distribution. The horizontal black error bars show the gravitational-wave binary black hole observations from O1 and O2 \citep{Collaboration2018GWTC-1:Runsa,Collaboration2018BinaryVirgoa}; their vertical positions are arbitrary.
}
\label{fig:BBH_chirp_mass_observed}
\end{figure*} 

\section{Observed binary black hole mass distribution}
\label{sec:with_selection_effects}

GW observations have strong selection effects favouring more massive systems; the maximum distance a BBH merger is observable to is approximately proportional to $\mathcal{M}^{5/6}$ (for low masses), meaning the observable volume is proportional to $\mathcal{M}^{5/2}$ \citep{Stevenson:2015bqa}. This means that the observed distribution is expected to over-emphasize the high-mass tail, allowing it to be explored more readily.

Our method for incorporating GW selection effects is described in \citet{Barrett:2017fcw}. We assume a representative strain sensitivity for the aLIGO detectors during O1 and O2\footnote{We assume each detector has a strain sensitivity equivalent to that of the Hanford LIGO detector during the observation of GW150914, publicly available at \url{https://dcc.ligo.org/LIGO-P1500238/public} \citep{Abbott:2016blz,TheLIGOScientific:2016zmo}. We have verified that assuming a different representative sensitivity from O2, publicly available at \url{https://dcc.ligo.org/LIGO-P1900011/public}, does not significantly change our results.}. We assume a single interferometer detection threshold signal-to-noise ratio of 8. We calculate the predicted observed rate as
\begin{equation}
    \frac{\mathrm{d}^2 N_\mathrm{BBH}}{\mathrm{d} t_\mathrm{obs} \mathrm{d} z} = \frac{\mathrm{d}^2 N_\mathrm{BBH}}{\mathrm{d} t_\mathrm{s} \mathrm{d} V_\mathrm{c}} \frac{\mathrm{d} V_\mathrm{c}}{\mathrm{d} z}
    \frac{\mathrm{d} t_\mathrm{s}}{\mathrm{d} t_\mathrm{obs}}
    P_\mathrm{det} ,
    \label{eq:with_selection_effects}
\end{equation}
where the first term is given by Equation~\ref{eq:rateeq}, the second term is the differential comoving volume at redshift $z$ \citep[e.g.][]{Hogg:1999ad}\footnote{We use \texttt{astropy}'s \texttt{differential\_comoving\_volume} function.}, $t_\mathrm{s}$ is time in the source frame and $t_\mathrm{obs} = (1 + z) t_\mathrm{s}$ is time in the observer’s frame. $P_\mathrm{det}$ is the detection probability for a binary with a given set of masses and a given distance, averaged over sky locations and source orientations.

We show the predicted observed redshift distribution in Figure~\ref{fig:BBH_redshift_distributions}. We see that in our model most BBH mergers during O1 and O2 would have been expected to be observed at redshift $z \sim 0.1$. 

Our models predict a total observed rate of $\sim 20$ BBHs per year of observing time, integrated over redshift. Since O1 and O2 combined constitute around 0.5 yr of observing time \citep{Collaboration2018GWTC-1:Runsa}, our model predicts $\sim 10$ BBH detections during O1 and O2, in agreement with observations. 


In Figure~\ref{fig:BBH_larger_mass_observed} we show the predicted observed distribution of the more massive black hole. We define $\Lambda_\mathrm{obs}$ as the fraction of observed BBHs in which at least one of the BHs was formed from a progenitor which experienced a PPISN, similarly to Section~\ref{sec:results}. We find that $\Lambda_\mathrm{obs} = 0.2$--$0.5$ in our models (see Table~\ref{table:results}). This suggests that we would expect $2$--$5$ of the observed BBH mergers to be such systems, strengthening our hypothesis that the more massive black holes in both GW150914 and GW170729 may have formed this way. 

We also show the predicted observed chirp mass distributions for our models in Figure~\ref{fig:BBH_chirp_mass_observed}. \citet{Marchant:2018kun} use their models, along with a toy model for the BH mass distribution, to investigate the expected observed BH chirp mass distribution. Accounting for GW selection effects, their model predicts that the distribution of observed chirp masses will show a double peak structure. We do not see such a feature in our chirp mass distributions shown in Figure~\ref{fig:BBH_chirp_mass_observed}.


\section{Discussion and Caveats}
\label{sec:discussion}

\subsection{Results Interpretation}
\label{subsec:results_interpretation}

BBHs merging at redshift $z=0$ where at least one BH formed from a progenitor that underwent a PPISN are intrinsically rare, with a volumetric rate of $\sim 3$~\PerCubicGigaparsecPerYear (see Figure~\ref{fig:rate_v_redshift}). Meanwhile, the merger rate of massive BBHs has been empirically determined through GW observations to be $\sim 10$--$40$~\PerCubicGigaparsecPerYear{} \citep{Collaboration2018BinaryVirgoa}. In addition to the absolute rate, we also present the fraction $\Lambda_\mathrm{merge}$ of BBHs in which at least one BH formed from a progenitor that underwent a PPISN. In our models, $\Lambda_\mathrm{merge} \sim 0.1$ at redshift $z=0$, meaning that around 10\% of merging BBHs form in this way (see Figure~\ref{fig:proportion_ppisn}). This value is towards the low end of the empirically estimated value $\Lambda_\mathrm{LVC} = 0.4^{+0.3}_{-0.3}$ \citep{Collaboration2018BinaryVirgoa}.

Our models suggest that it is unlikely that BBHs formed from progenitors which underwent PPISNe can contribute more significantly than this to the population merging at redshift $z = 0$. We suggest that an alternative explanation may be required, such as a model for the distribution of the mass of the more massive black hole that is not a pure power law across the entire mass range.

Although BBHs merging at redshift $z = 0$ with at least one component formed from a progenitor which underwent PPISNe are intrinsically rare, GW observations strongly favor massive systems. Our models predict that $20$--$50$\% of observed BBHs will have at least one component formed from a progenitor that underwent PPISNe (see e.g. Figure~\ref{fig:BBH_larger_mass_observed}). This lends support to the hypothesis that some of the observed BBH systems may have formed this way.

We have also calculated the volumetric rates of CCSNe, PISNe and PPISNe predicted by our models as a function of redshift (see Figure~\ref{fig:rate_v_redshift}). The rate of both PISNe and PPISNe track the low metallicity star formation rate. The volumetric rate of CCSNe (including the stars which undergo PPISNe) at redshift $z=0$ in our model is $\sim 2 \times 10^4$~\PerCubicGigaparsecPerYear{}, whilst the rate of PISNe is $\lesssim 10^{-2}$~\PerCubicGigaparsecPerYear{}. We find the rate of stars undergoing PPISNe to be $\sim 0.1$~\PerCubicGigaparsecPerYear{}. Our rates for CCSNe, PPISNe and PISNe should be treated as lower limits from our model populations, since there are evolutionary channels we have not included, such as supernovae from stellar merger products \citep{Vigna-Gomez:2019sky} and supernovae in unbound systems. Our CCSNe rate includes those events where a BH forms through complete fallback; such an event may lead to a `failed' (and therefore not observed) supernova \citep{Gerke:2015MNRAS,Adams:2016ffj}.

In addition, our rate of systems undergoing PPISNe is likely lower than the actual rate of observed PPISNe, since each star may undergo multiple widely-spaced PPISN events before collapsing \citep{Woosley:2007qp,Woosley:2016hmi,Marchant:2018kun}. 

PISNe, PPISNe and BBH mergers are inherently rare events in our populations. This leads to large statistical uncertainties on some of our predictions (see Table~\ref{table:results}). \citet{Broekgaarden:2019qnw} have recently shown how binary population synthesis can much more effectively explore rare populations of astrophysical phenomena. These techniques can be applied to study BBH mergers, PISNe and PPISNe.

\subsection{Impact of metallicity specific star formation rate}
\label{subsec:impact_of_MSSFR}

The BBH merger rate tracks the low metallicity star formation rate, increasing from redshift $z = 0$ to $z = 2$. The fraction of star formation occurring at low metallicities is poorly constrained observationally \citep[e.g.][]{Madau:2014bja}, particularly at the high redshifts ($z > 2$) where BBHs may form in large numbers. 


\citet{coen} show that uncertainties in the metallicity specific star formation rate can change predictions for the BBH merger rate by an order of magnitude \citep[see also][]{Chruslinska:2018hrb}, comparable to uncertainties in binary evolution \citep{Dominik:2012kk}. Since the occurrence of PISNe and PPISNe depends strongly on metallicity (see Figure~\ref{fig:formation_efficiency_per_solar_mass_star_formation}), these uncertainties also have a large impact on our predictions of these events. For example, in combination with our default binary evolution assumptions, using the star formation rate from \citet{Madau:2014bja} and the metallicity distribution from \citet{Langer:2005hu} leads to $\Lambda_\mathrm{merge}$ being a factor of $10$ lower than in our default model. This primarily arises due to an increase in the merger rate of low mass ($\mathcal{M} < 10 M_\odot$) BBH mergers. This metallicity-specific star formation history model over-predicts the total BBH merger rate \citep{coen}, compared to GW observations \citet{Collaboration2018BinaryVirgoa}. 

\subsection{Binary evolution}
\label{subsec:binary_evolution_caveats}

There are also significant uncertainties in massive binary evolution. Previous studies have shown that uncertainties in black hole kicks, the initial mass function, stellar wind mass loss, common envelope evolution and mass transfer, may lead to up to an order of magnitude uncertainty in BBH merger rates \cite[e.g.][]{Dominik:2012kk}. In the models we have presented here, we have only investigated uncertainties in our modelling of PISNe and PPISNe. We have not attempted to quantify the uncertainies in the model predictions due to uncertain physics. We will investigate what constraints current GW observations place on binary evolution in the future.

\subsection{Stellar models}
\label{subsec:stellar_model_uncertainty}

Another source of uncertainty inherent in our population synthesis method is the reliance on pre-computed stellar evolutionary models. The stellar models implemented in COMPAS \citep{Pols:1998MNRAS,Hurley:2000pk} were originally computed for the mass range of $5$--$50$~\Msol{}. This means that we needed to extrapolate up to $150$\,\Msol{} for the present study. Such an extrapolation, while standard practice in population synthesis \citep[e.g.][]{Dominik:2012kk,coen}, may need to be revisited and possibly refined in future work. Additionally, there have been significant updates in massive stellar evolution since the models we rely on were published.  For example, the role of rotation has been extensively investigated and shown to be important in driving the evolution of a star, especially at low metallicities
\citep[cf.][]{Brott:2011A&A,Szecsi:2015A&A}. Another recently studied effect that may influence our results is the possibility of `envelope inflation' in stars with $\gtrsim$~70~M$_{\odot}$ \citep{Sanyal:2015A&A,Sanyal:2017A&A}. This may lead to differences in when and how a binary interacts and thus to differences in the final outcome of our population synthesis code. The update of our code in order to be able to interpolate on-the-fly from pre-computed stellar tracks \citep[cf.][]{Kruckow:2018slo,Spera:2018wnw} in order to reflect these recent developments in massive stellar evolution is currently under way.


\subsection{Literature comparison}
\label{subsec:literature_comparison}

We now turn to comparing our predicted event rates to those in the literature. \citet{Eldridge:2018nop} assume the star formation rate from \citet{Madau:2014bja} and the metallicity distribution form \citet{Langer:2005hu} and find the volumetric rate of CCSNe at redshift $z = 0$ in their models to be $\sim 10^{5}$~\PerCubicGigaparsecPerYear, whilst they predict the PISN rate to be $\sim 10$~\PerCubicGigaparsecPerYear, and their BBH merger rate is $\sim 10^2$~\PerCubicGigaparsecPerYear{} at the same redshift. They do not provide a rate for PPISNe. \citet{Eldridge:2018RNAAS} also provide chirp mass distributions using a simplified method of accounting for GW selection effects. 


\citet{Nicholl:2013Nature} argue that based on a lack of unambiguous observations of local PISNe, the local PISN rate must be less than $6 \times 10^{-6}$ of the local CCSN rate. This is compatible with our predicted rates at redshift $z=0$ which give $\mathcal{R}_\mathrm{PISN} / \mathcal{R}_\mathrm{CCSN} \lesssim 10^{-6}$. \citet{Eldridge:2018nop} find $\mathcal{R}_\mathrm{PISN} / \mathcal{R}_\mathrm{CCSN} \sim 10^{-4}$ in their model.

PISNe and PPISNe may account for a small fraction of the observed SLSNe \citep{2015MNRAS.454.4357K,2018ApJ...858..115A}. SLSNe are observable to cosmological distances of redshifts $z \sim 3$--$4$ \citep[e.g.][]{Cooke:2012Nature,Moryia:2018arXiv180108240M}. The rate of SLSNe inferred from observations at redshift $z \sim 0$ is $\mathcal{R}_\mathrm{SLSNe} \sim 100$~\PerCubicGigaparsecPerYear, rising to $\mathcal{R}_\mathrm{SLSNe} \sim 1000$~\PerCubicGigaparsecPerYear at $z = 2$ before falling again to $\mathcal{R}_\mathrm{SLSNe} \sim 500$~\PerCubicGigaparsecPerYear at $z = 3$--$4$ \citep{Quimby:2013jb,Cooke:2012Nature,Prajs:2016cjj,Moryia:2018arXiv180108240M}. Since it is currently unclear what fraction of SLSNe are due to PISNe and PPISNe, a direct comparison is not possible. 

\citet{Marchant:2018kun} suggest that in close binaries which go on to form BBHs, the expansion of the star during a PPISN may lead to an additional period of mass transfer, or even common envelope evolution. We have neglected this effect in this work. 

\section{Conclusion}
\label{sec:conclusion}

In this paper we have investigated the effect of PPISNe and PISNe on the BBH mass distribution, which is being unveiled through GW observations \citep{Collaboration2018GWTC-1:Runsa,Collaboration2018BinaryVirgoa}. These observations are providing evidence that the maximum mass of BHs in merging BBHs is $\sim 45$~\Msol{}, and are hinting at an excess of BHs in the mass range $30$--$45$\,\Msol{} \citep{Collaboration2018GWTC-1:Runsa,Collaboration2018BinaryVirgoa}. In this paper, we have investigated whether it is plausible to attribute these features to PISNe and PPISNe in massive binaries.

We use simple analytic fits to more detailed models to describe the relation between pre-supernova helium core mass and final remnant mass \citep{Belczynski:2016jno,Woosley:2016hmi,Marchant:2018kun,Woosley2019TheLoss} coupled with the population synthesis code COMPAS \citep{Stevenson2017FormationEvolution,2018MNRAS.481.4009V,Barrett:2017fcw,Broekgaarden:2019qnw,coen,bavera}. 

Our models predict a maximum BH mass of $40$--$50$\,\Msol{}, in agreement with GW observations. More massive pre-supernova helium cores undergo PISNe and are completely unbound, leaving no remnant. BHs in the mass range $35$--$45$\,\Msol{} form exclusively from progenitors which underwent PPISNe in our models (except in the model of \citet{Belczynski:2016jno}), and can be readily identified from GW observations. However, GW measurement uncertainties for such massive BBH mergers are typically $> 10$\,\Msol{} \citep[see e.g.][]{Collaboration2018GWTC-1:Runsa}, hindering definite association in many cases. We suggest that the more massive black hole in the BBH merger GW150914 \citep{Abbott:2016blz} may have formed from a progenitor which underwent PPISNe, along with potentially both components of the BBH merger GW170729 if the ratio of the mass of the two black holes is close to 1 \citep[][]{Collaboration2018GWTC-1:Runsa}.

We find that PPISNe lead to a more gradual transition into the PISN mass gap, which will complicate efforts to determine the exact mass coordinate of the gap. Sharp cutoffs can be measured with an accuracy that scales as $1/N$, not $1/\sqrt{N}$ \citep[e.g.][]{Mandel:2014CQG}. 

As ground-based GW detectors continue to increase in sensitivity, the number of observations of BBHs will continue to grow \citep{Aasi:2013wya}. Their increased reach will enable measurements of the BBH merger rate as a function of redshift \citep{Fishbach:2018edt}, which can then be compared to the models presented in this paper. A subpopulation of BBHs formed from progenitors which underwent PPISNe may be uncovered. This will allow us to study these types of supernovae indirectly using GWs, increasing our understanding of massive stellar evolution.

In this study, we have used the 10 BBH observations from LIGO and Virgo, published in the GWTC-1 catalog \citep{Collaboration2018GWTC-1:Runsa}. Recently, \citet{Zackay:2019tzo} and \citet{Venumadhav:2019lyq} have almost doubled the number of BBH candidates observed during O1 and O2 using different data analysis techniques. Future comparisons to GW observations should include all observations in order to avoid biased astrophysical conclusions.

Additionally, in this paper we have focused on the mass distribution of BBHs; we present our predictions for the spin distribution of binary black holes in \citet{bavera}.

\section*{Acknowledgements}

We thank Bernhard Müller, Eric Thrane and Alexandra Kozyreva for useful discussions and comments on this paper. We also thank the anonymous referee for insightful comments and suggestions. The authors are supported by the Australian Research Council Centre of Excellence for Gravitational Wave Discovery (OzGrav), through project number CE170100004. This work made use of the OzSTAR high performance computer at Swinburne University of Technology. OzSTAR is funded by Swinburne University of Technology and the National Collaborative Research Infrastructure Strategy (NCRIS). D.Sz. accepts funding from the Alexander von Humboldt Foundation. AVG acknowledges funding support from CONACYT. This research made use of Astropy,\footnote{http://www.astropy.org} a community-developed core Python package for Astronomy \citep{astropy:2013, astropy:2018}.

\bibliographystyle{apj}
\bibliography{bib.bib} 

\begin{thebibliography}{}
\expandafter\ifx\csname natexlab\endcsname\relax\def\natexlab#1{#1}\fi

\bibitem[{Aasi {et~al.}(2015)}]{TheLIGOScientific:2014jea}
Aasi, J., {et~al.} 2015, Class. Quant. Grav., 32, 074001

\bibitem[{Abbott {et~al.}(2017)}]{TheLIGOScientific:2017qsa}
Abbott, B., {et~al.} 2017, Phys. Rev. Lett., 119, 161101

\bibitem[{Abbott {et~al.}(2016{\natexlab{a}})}]{TheLIGOScientific:2016zmo}
Abbott, B.~P., {et~al.} 2016{\natexlab{a}}, Class. Quant. Grav., 33, 134001

\bibitem[{Abbott {et~al.}(2016{\natexlab{b}})}]{Abbott:2016blz}
---. 2016{\natexlab{b}}, Phys. Rev. Lett., 116, 061102

\bibitem[{Abbott {et~al.}(2018)}]{Aasi:2013wya}
---. 2018, Living Rev. Rel., 21, 3

\bibitem[{Acernese {et~al.}(2015)}]{TheVirgo:2014hva}
Acernese, F., {et~al.} 2015, Class. Quant. Grav., 32, 024001

\bibitem[{Adams {et~al.}(2017)Adams, Kochanek, Gerke, Stanek, \&
  Dai}]{Adams:2016ffj}
Adams, S.~M., Kochanek, C.~S., Gerke, J.~R., Stanek, K.~Z., \& Dai, X. 2017,
  Mon. Not. Roy. Astron. Soc., 468, 4968

\bibitem[{{Aguilera-Dena} {et~al.}(2018){Aguilera-Dena}, {Langer}, {Moriya}, \&
  {Schootemeijer}}]{2018ApJ...858..115A}
{Aguilera-Dena}, D.~R., {Langer}, N., {Moriya}, T.~J., \& {Schootemeijer}, A.
  2018, \apj, 858, 115

\bibitem[{Antonini {et~al.}(2017)Antonini, Toonen, \&
  Hamers}]{Antonini:2017ash}
Antonini, F., Toonen, S., \& Hamers, A.~S. 2017, Astrophys. J., 841, 77

\bibitem[{{Arcavi} {et~al.}(2017){Arcavi}, {Howell}, {Kasen}, {Bildsten},
  {Hosseinzadeh}, {McCully}, {Wong}, {Katz}, {Gal-Yam}, {Sollerman}, {Taddia},
  {et~al.}}]{Arcavi2017Nature}
{Arcavi}, I., {Howell}, D.~A., {Kasen}, D., {et~al.} 2017, \nat, 551, 210

\bibitem[{{Astropy Collaboration} {et~al.}(2013){Astropy Collaboration},
  {Robitaille}, {Tollerud}, {Greenfield}, {Droettboom}, {Bray}, {Aldcroft},
  {Davis}, {Ginsburg}, {Price-Whelan}, {Kerzendorf}, {Conley}, {Crighton},
  {Barbary}, {Muna}, {Ferguson}, {Grollier}, {Parikh}, {Nair}, {Unther},
  {Deil}, {Woillez}, {Conseil}, {Kramer}, {Turner}, {Singer}, {Fox}, {Weaver},
  {Zabalza}, {Edwards}, {Azalee Bostroem}, {Burke}, {Casey}, {Crawford},
  {Dencheva}, {Ely}, {Jenness}, {Labrie}, {Lim}, {Pierfederici}, {Pontzen},
  {Ptak}, {Refsdal}, {Servillat}, \& {Streicher}}]{astropy:2013}
{Astropy Collaboration}, {Robitaille}, T.~P., {Tollerud}, E.~J., {et~al.} 2013,
  \aap, 558, A33

\bibitem[{{Barkat} {et~al.}(1967){Barkat}, {Rakavy}, \&
  {Sack}}]{1967PhRvL..18..379B}
{Barkat}, Z., {Rakavy}, G., \& {Sack}, N. 1967, Physical Review Letters, 18,
  379

\bibitem[{Barrett {et~al.}(2018)Barrett, Gaebel, Neijssel, Vigna-Gómez,
  Stevenson, Berry, Farr, \& Mandel}]{Barrett:2017fcw}
Barrett, J.~W., Gaebel, S.~M., Neijssel, C.~J., {et~al.} 2018, Mon. Not. Roy.
  Astron. Soc., 477, 4685

\bibitem[{Bartos {et~al.}(2017)Bartos, Kocsis, Haiman, \&
  Márka}]{Bartos:2016dgn}
Bartos, I., Kocsis, B., Haiman, Z., \& Márka, S. 2017, Astrophys. J., 835, 165

\bibitem[{{Bavera} {et~al.}(2019){Bavera}, {Fragos}, {Qin}, {Zapartas},
  {Neijssel}, {Mandel}, {Batta}, {Gaebel}, {Kimball}, \& {Stevenson}}]{bavera}
{Bavera}, S.~S., {Fragos}, T., {Qin}, Y., {et~al.} 2019, arXiv e-prints,
  arXiv:1906.12257

\bibitem[{{Belczynski} {et~al.}(2010){Belczynski}, {Bulik}, {Fryer}, {Ruiter},
  {Valsecchi}, {Vink}, \& {Hurley}}]{Belczynski:2010ApJ}
{Belczynski}, K., {Bulik}, T., {Fryer}, C.~L., {et~al.} 2010, \apj, 714, 1217

\bibitem[{Belczynski {et~al.}(2016{\natexlab{a}})Belczynski, Holz, Bulik, \&
  O'Shaughnessy}]{Belczynski:2016obo}
Belczynski, K., Holz, D.~E., Bulik, T., \& O'Shaughnessy, R.
  2016{\natexlab{a}}, Nature, 534, 512

\bibitem[{Belczynski {et~al.}(2007)Belczynski, Kalogera, Rasio, Taam, \&
  Bulik}]{Belczynski:2006zi}
Belczynski, K., Kalogera, V., Rasio, F.~A., Taam, R.~E., \& Bulik, T. 2007,
  Astrophys. J., 662, 504

\bibitem[{Belczynski {et~al.}(2016{\natexlab{b}})}]{Belczynski:2016jno}
Belczynski, K., {et~al.} 2016{\natexlab{b}}, Astron. Astrophys., 594, A97

\bibitem[{Bird {et~al.}(2016)Bird, Cholis, Muñoz, Ali-Haïmoud, Kamionkowski,
  Kovetz, Raccanelli, \& Riess}]{Bird:2016dcv}
Bird, S., Cholis, I., Muñoz, J.~B., {et~al.} 2016, Phys. Rev. Lett., 116,
  201301

\bibitem[{{Blaauw}(1961)}]{1961BAN....15..265B}
{Blaauw}, A. 1961, \bain, 15, 265

\bibitem[{{Boersma}(1961)}]{1961BAN....15..291B}
{Boersma}, J. 1961, \bain, 15, 291

\bibitem[{{Broadhurst} {et~al.}(2019){Broadhurst}, {Diego}, \&
  {Smoot}}]{Broadhurst:2019ijv}
{Broadhurst}, T., {Diego}, J.~M., \& {Smoot}, George~F., I. 2019, arXiv
  e-prints, arXiv:1901.03190

\bibitem[{{Broekgaarden} {et~al.}(2019){Broekgaarden}, {Justham}, {de Mink},
  {Gair}, {Mandel}, {Stevenson}, {Barrett}, {Vigna-G{\'o}mez}, \&
  {Neijssel}}]{Broekgaarden:2019qnw}
{Broekgaarden}, F.~S., {Justham}, S., {de Mink}, S.~E., {et~al.} 2019, arXiv
  e-prints, arXiv:1905.00910

\bibitem[{{Brott} {et~al.}(2011){Brott}, {de Mink}, {Cantiello}, {Langer}, {de
  Koter}, {Evans}, {Hunter}, {Trundle}, \& {Vink}}]{Brott:2011A&A}
{Brott}, I., {de Mink}, S.~E., {Cantiello}, M., {et~al.} 2011, \aap, 530, A115

\bibitem[{{Castor} {et~al.}(1975){Castor}, {Abbott}, \&
  {Klein}}]{1975ApJ...195..157C}
{Castor}, J.~I., {Abbott}, D.~C., \& {Klein}, R.~I. 1975, \apj, 195, 157

\bibitem[{Chan {et~al.}(2018)Chan, Müller, Heger, Pakmor, \&
  Springel}]{Chan:2017tdg}
Chan, C., Müller, B., Heger, A., Pakmor, R., \& Springel, V. 2018, Astrophys.
  J., 852, L19

\bibitem[{Chatterjee {et~al.}(2017)Chatterjee, Rodriguez, Kalogera, \&
  Rasio}]{Chatterjee:2016thb}
Chatterjee, S., Rodriguez, C.~L., Kalogera, V., \& Rasio, F.~A. 2017,
  Astrophys. J., 836, L26

\bibitem[{{Chatzopoulos} \& {Wheeler}(2012)}]{Chatzopoulos:2012ApJ}
{Chatzopoulos}, E., \& {Wheeler}, J.~C. 2012, \apj, 748, 42

\bibitem[{Chruslinska {et~al.}(2019)Chruslinska, Nelemans, \&
  Belczynski}]{Chruslinska:2018hrb}
Chruslinska, M., Nelemans, G., \& Belczynski, K. 2019, Mon. Not. Roy. Astron.
  Soc., 482, 5012

\bibitem[{{Cooke} {et~al.}(2012){Cooke}, {Sullivan}, {Gal-Yam}, {Barton},
  {Carlberg}, {Ryan-Weber}, {Horst}, {Omori}, \&
  {D{\'\i}az}}]{Cooke:2012Nature}
{Cooke}, J., {Sullivan}, M., {Gal-Yam}, A., {et~al.} 2012, \nat, 491, 228

\bibitem[{{de Kool}(1990)}]{1990ApJ...358..189D}
{de Kool}, M. 1990, \apj, 358, 189

\bibitem[{de~Mink \& Belczynski(2015)}]{deMink:2015yea}
de~Mink, S.~E., \& Belczynski, K. 2015, Astrophys. J., 814, 58

\bibitem[{Di~Carlo {et~al.}(2019)Di~Carlo, Giacobbo, Mapelli, Pasquato, Spera,
  Wang, \& Haardt}]{DiCarlo:2019pmf}
Di~Carlo, U.~N., Giacobbo, N., Mapelli, M., {et~al.} 2019, arXiv e-prints,
  arXiv:1901.00863

\bibitem[{Dominik {et~al.}(2012)Dominik, Belczynski, Fryer, Holz, Berti, Bulik,
  Mandel, \& O'Shaughnessy}]{Dominik:2012kk}
Dominik, M., Belczynski, K., Fryer, C., {et~al.} 2012, Astrophys. J., 759, 52

\bibitem[{{Eggleton}(1983)}]{Eggleton:1983ApJ}
{Eggleton}, P.~P. 1983, \apj, 268, 368

\bibitem[{Eldridge \& Stanway(2016)}]{Eldridge:2016ymr}
Eldridge, J.~J., \& Stanway, E.~R. 2016, Mon. Not. Roy. Astron. Soc., 462, 3302

\bibitem[{{Eldridge} {et~al.}(2019){Eldridge}, {Stanway}, \&
  {Tang}}]{Eldridge:2018nop}
{Eldridge}, J.~J., {Stanway}, E.~R., \& {Tang}, P.~N. 2019, \mnras, 482, 870

\bibitem[{Eldridge {et~al.}(2018)Eldridge, Tang, Bray, \&
  Stanway}]{Eldridge:2018RNAAS}
Eldridge, J.~J., Tang, P., Bray, J., \& Stanway, E.~R. 2018, Research Notes of
  the {AAS}, 2, 236

\bibitem[{{Farr} \& {Mandel}(2018)}]{FarrMandel:2018}
{Farr}, W.~M., \& {Mandel}, I. 2018, Science, 361, aat6506

\bibitem[{Fishbach \& Holz(2017)}]{Fishbach:2017zga}
Fishbach, M., \& Holz, D.~E. 2017, Astrophys. J., 851, L25

\bibitem[{{Fishbach} {et~al.}(2017){Fishbach}, {Holz}, \&
  {Farr}}]{2017ApJ...840L..24F}
{Fishbach}, M., {Holz}, D.~E., \& {Farr}, B. 2017, \apjl, 840, L24

\bibitem[{Fishbach {et~al.}(2018)Fishbach, Holz, \& Farr}]{Fishbach:2018edt}
Fishbach, M., Holz, D.~E., \& Farr, W.~M. 2018, Astrophys. J., 863, L41,
  [Astrophys. J. Lett.863,L41(2018)]

\bibitem[{{Fowler} \& {Hoyle}(1964)}]{FowlerHoyle:1964ApJS}
{Fowler}, W.~A., \& {Hoyle}, F. 1964, \apjs, 9, 201

\bibitem[{Fragione \& Kocsis(2018)}]{Fragione:2018vty}
Fragione, G., \& Kocsis, B. 2018, Phys. Rev. Lett., 121, 161103

\bibitem[{Fragione \& Kocsis(2019)}]{Fragione:2019hqt}
---. 2019, arXiv e-prints, arXiv:1903.03112

\bibitem[{{Fraley}(1968)}]{1968Ap&SS...2...96F}
{Fraley}, G.~S. 1968, \apss, 2, 96

\bibitem[{{Fryer} {et~al.}(2012){Fryer}, {Belczynski}, {Wiktorowicz},
  {Dominik}, {Kalogera}, \& {Holz}}]{Fryer:2012ApJ}
{Fryer}, C.~L., {Belczynski}, K., {Wiktorowicz}, G., {et~al.} 2012, \apj, 749,
  91

\bibitem[{{Fryer} {et~al.}(2001){Fryer}, {Woosley}, \& {Heger}}]{Fryer:2001ApJ}
{Fryer}, C.~L., {Woosley}, S.~E., \& {Heger}, A. 2001, \apj, 550, 372

\bibitem[{{Gal-Yam}(2012)}]{2012Sci...337..927G}
{Gal-Yam}, A. 2012, Science, 337, 927

\bibitem[{{Gal-Yam} {et~al.}(2009){Gal-Yam}, {Mazzali}, {Ofek}, {Nugent},
  {Kulkarni}, {Kasliwal}, {Quimby}, {Filippenko}, {Cenko}, {Chornock},
  {Waldman}, {Kasen}, {Sullivan}, {Beshore}, {Drake}, {Thomas}, {Bloom},
  {Poznanski}, {Miller}, {Foley}, {Silverman}, {Arcavi}, {Ellis}, \&
  {Deng}}]{GalYam:2009Nature}
{Gal-Yam}, A., {Mazzali}, P., {Ofek}, E.~O., {et~al.} 2009, \nat, 462, 624

\bibitem[{{Gerke} {et~al.}(2015){Gerke}, {Kochanek}, \&
  {Stanek}}]{Gerke:2015MNRAS}
{Gerke}, J.~R., {Kochanek}, C.~S., \& {Stanek}, K.~Z. 2015, \mnras, 450, 3289

\bibitem[{Gerosa \& Berti(2017)}]{Gerosa:2017kvu}
Gerosa, D., \& Berti, E. 2017, Phys. Rev., D95, 124046

\bibitem[{Giacobbo {et~al.}(2018)Giacobbo, Mapelli, \&
  Spera}]{Giacobbo:2017qhh}
Giacobbo, N., Mapelli, M., \& Spera, M. 2018, Mon. Not. Roy. Astron. Soc., 474,
  2959

\bibitem[{Gomez {et~al.}(2019)Gomez, Berger, Nicholl, Blanchard, Villar,
  Patton, Chornock, Leja, Hosseinzadeh, \& Cowperthwaite}]{Gomez:2019yeb}
Gomez, S., Berger, E., Nicholl, M., {et~al.} 2019, arXiv e-prints,
  arXiv:1904.07259

\bibitem[{{Hayashi} \& {Nakano}(1963)}]{1963PThPh..30..460H}
{Hayashi}, C., \& {Nakano}, T. 1963, Progress of Theoretical Physics, 30, 460

\bibitem[{{Hinshaw} {et~al.}(2013){Hinshaw}, {Larson}, {Komatsu}, {Spergel},
  {Bennett}, {Dunkley}, {Nolta}, {Halpern}, {Hill}, {Odegard}, {Page}, {Smith},
  {Weiland}, {Gold}, {Jarosik}, {Kogut}, {Limon}, {Meyer}, {Tucker}, {Wollack},
  \& {Wright}}]{2013ApJS..208...19H}
{Hinshaw}, G., {Larson}, D., {Komatsu}, E., {et~al.} 2013, \apjs, 208, 19

\bibitem[{{Hjellming} \& {Webbink}(1987)}]{1987HjellmingRadii}
{Hjellming}, M.~S., \& {Webbink}, R.~F. 1987, \apj, 318, 794

\bibitem[{Hobbs {et~al.}(2005)Hobbs, Lorimer, Lyne, \& Kramer}]{Hobbs:2005yx}
Hobbs, G., Lorimer, D.~R., Lyne, A.~G., \& Kramer, M. 2005, Mon. Not. Roy.
  Astron. Soc., 360, 974

\bibitem[{Hogg(1999)}]{Hogg:1999ad}
Hogg, D.~W. 1999, arXiv e-prints, arXiv:astro-ph/9905116

\bibitem[{{Humphreys} \& {Davidson}(1994)}]{1994PASP..106.1025H}
{Humphreys}, R.~M., \& {Davidson}, K. 1994, \pasp, 106, 1025

\bibitem[{Hurley {et~al.}(2000)Hurley, Pols, \& Tout}]{Hurley:2000pk}
Hurley, J.~R., Pols, O.~R., \& Tout, C.~A. 2000, Mon. Not. Roy. Astron. Soc.,
  315, 543

\bibitem[{Hurley {et~al.}(2002)Hurley, Tout, \& Pols}]{Hurley:2002rf}
Hurley, J.~R., Tout, C.~A., \& Pols, O.~R. 2002, Mon. Not. Roy. Astron. Soc.,
  329, 897

\bibitem[{{Ivanova} {et~al.}(2013){Ivanova}, {Justham}, {Chen}, {De Marco},
  {Fryer}, {Gaburov}, {Ge}, {Glebbeek}, {Han}, {Li},
  {et~al.}}]{2013A&ARv..21...59I}
{Ivanova}, N., {Justham}, S., {Chen}, X., {et~al.} 2013, Astronomy and
  Astrophysics Review, 21, 59

\bibitem[{Janka(2013)}]{Janka:2013hfa}
Janka, H.~T. 2013, Mon. Not. Roy. Astron. Soc., 434, 1355

\bibitem[{Kimball {et~al.}(2019)Kimball, Berry, \& Kalogera}]{Kimball:2019mfs}
Kimball, C., Berry, C. P.~L., \& Kalogera, V. 2019, arXiv e-prints,
  arXiv:1903.07813

\bibitem[{Klencki {et~al.}(2018)Klencki, Moe, Gladysz, Chruslinska, Holz, \&
  Belczynski}]{Klencki:2018zrz}
Klencki, J., Moe, M., Gladysz, W., {et~al.} 2018, Astron. Astrophys., 619, A77

\bibitem[{{Kovetz} {et~al.}(2017){Kovetz}, {Cholis}, {Breysse}, \&
  {Kamionkowski}}]{2017PhRvD..95j3010K}
{Kovetz}, E.~D., {Cholis}, I., {Breysse}, P.~C., \& {Kamionkowski}, M. 2017,
  \prd, 95, 103010

\bibitem[{{Kozyreva} \& {Blinnikov}(2015)}]{2015MNRAS.454.4357K}
{Kozyreva}, A., \& {Blinnikov}, S. 2015, \mnras, 454, 4357

\bibitem[{{Kozyreva} {et~al.}(2014{\natexlab{a}}){Kozyreva}, {Blinnikov},
  {Langer}, \& {Yoon}}]{2014A&A...565A..70K}
{Kozyreva}, A., {Blinnikov}, S., {Langer}, N., \& {Yoon}, S.-C.
  2014{\natexlab{a}}, \aap, 565, A70

\bibitem[{{Kozyreva} {et~al.}(2018){Kozyreva}, {Kromer}, {Noebauer}, \&
  {Hirschi}}]{Kozyreva:2018MNRAS}
{Kozyreva}, A., {Kromer}, M., {Noebauer}, U.~M., \& {Hirschi}, R. 2018, \mnras,
  479, 3106

\bibitem[{{Kozyreva} {et~al.}(2014{\natexlab{b}}){Kozyreva}, {Yoon}, \&
  {Langer}}]{2014A&A...566A.146K}
{Kozyreva}, A., {Yoon}, S.-C., \& {Langer}, N. 2014{\natexlab{b}}, \aap, 566,
  A146

\bibitem[{{Kozyreva} {et~al.}(2017){Kozyreva}, {Gilmer}, {Hirschi},
  {Fr{\"o}hlich}, {Blinnikov}, {Wollaeger}, {Noebauer}, {van Rossum}, {Heger},
  {Even}, {Waldman}, {Tolstov}, {Chatzopoulos}, \&
  {Sorokina}}]{2017MNRAS.464.2854K}
{Kozyreva}, A., {Gilmer}, M., {Hirschi}, R., {et~al.} 2017, \mnras, 464, 2854

\bibitem[{Kroupa(2001)}]{Kroupa:2000iv}
Kroupa, P. 2001, Mon. Not. Roy. Astron. Soc., 322, 231

\bibitem[{{Kruckow} {et~al.}(2018){Kruckow}, {Tauris}, {Langer}, {Kramer}, \&
  {Izzard}}]{Kruckow:2018slo}
{Kruckow}, M.~U., {Tauris}, T.~M., {Langer}, N., {Kramer}, M., \& {Izzard},
  R.~G. 2018, \mnras, 481, 1908

\bibitem[{{Kulkarni} {et~al.}(1993){Kulkarni}, {Hut}, \&
  {McMillan}}]{1993Natur.364..421K}
{Kulkarni}, S.~R., {Hut}, P., \& {McMillan}, S. 1993, \nat, 364, 421

\bibitem[{{Kumar}(1963)}]{1963ApJ...137.1121K}
{Kumar}, S.~S. 1963, \apj, 137, 1121

\bibitem[{Langer \& Norman(2006)}]{Langer:2005hu}
Langer, N., \& Norman, C.~A. 2006, Astrophys. J., 638, L63

\bibitem[{Leung {et~al.}(2019)Leung, Nomoto, \& Blinnikov}]{Leung:2019fgj}
Leung, S.-C., Nomoto, K., \& Blinnikov, S. 2019, arXiv e-prints,
  arXiv:1901.11136

\bibitem[{{Lunnan} {et~al.}(2016){Lunnan}, {Chornock}, {Berger},
  {Milisavljevic}, {Jones}, {Rest}, {Fong}, {Fransson}, {Margutti}, {Drout},
  {Blanchard}, {Challis}, {Cowperthwaite}, {Foley}, {Kirshner}, {Morrell},
  {Riess}, {Roth}, {Scolnic}, {Smartt}, {Smith}, {Villar}, {Chambers},
  {Draper}, {Huber}, {Kaiser}, {Kudritzki}, {Magnier}, {Metcalfe}, \&
  {Waters}}]{Lunnan:2016ApJ}
{Lunnan}, R., {Chornock}, R., {Berger}, E., {et~al.} 2016, \apj, 831, 144

\bibitem[{{Lunnan} {et~al.}(2018){Lunnan}, {Fransson}, {Vreeswijk}, {Woosley},
  {Leloudas}, {Perley}, {Quimby}, {Yan}, {Blagorodnova}, {Bue}, {Cenko}, {De
  Cia}, {Cook}, {Fremling}, {Gatkine}, {Gal-Yam}, {Kasliwal}, {Kulkarni},
  {Masci}, {Nugent}, {Nyholm}, {Rubin}, {Suzuki}, \&
  {Wozniak}}]{Lunnan:2018NatAstr}
{Lunnan}, R., {Fransson}, C., {Vreeswijk}, P.~M., {et~al.} 2018, Nature
  Astronomy, 2, 887

\bibitem[{Madau \& Dickinson(2014)}]{Madau:2014bja}
Madau, P., \& Dickinson, M. 2014, Ann. Rev. Astron. Astrophys., 52, 415

\bibitem[{Madau \& Fragos(2017)}]{Madau:2016jbv}
Madau, P., \& Fragos, T. 2017, Astrophys. J., 840, 39

\bibitem[{Mandel(2016)}]{Mandel:2015eta}
Mandel, I. 2016, Mon. Not. Roy. Astron. Soc., 456, 578

\bibitem[{{Mandel} {et~al.}(2014){Mandel}, {Berry}, {Ohme}, {Fairhurst}, \&
  {Farr}}]{Mandel:2014CQG}
{Mandel}, I., {Berry}, C.~P.~L., {Ohme}, F., {Fairhurst}, S., \& {Farr}, W.~M.
  2014, Classical and Quantum Gravity, 31, 155005

\bibitem[{Mandel \& de~Mink(2016)}]{Mandel:2015qlu}
Mandel, I., \& de~Mink, S.~E. 2016, Mon. Not. Roy. Astron. Soc., 458, 2634

\bibitem[{Mandel \& Farmer(2018)}]{Mandel:2018hfr}
Mandel, I., \& Farmer, A. 2018, arXiv e-prints, arXiv:1806.05820

\bibitem[{Marchant {et~al.}(2016)Marchant, Langer, Podsiadlowski, Tauris, \&
  Moriya}]{Marchant:2016wow}
Marchant, P., Langer, N., Podsiadlowski, P., Tauris, T.~M., \& Moriya, T.~J.
  2016, Astron. Astrophys., 588, A50

\bibitem[{{Marchant} {et~al.}(2018){Marchant}, {Renzo}, {Farmer}, {Pappas},
  {Taam}, {de Mink}, \& {Kalogera}}]{Marchant:2018kun}
{Marchant}, P., {Renzo}, M., {Farmer}, R., {et~al.} 2018, arXiv e-prints,
  arXiv:1810.13412

\bibitem[{Michaely \& Perets(2019)}]{Michaely:2019aet}
Michaely, E., \& Perets, H.~B. 2019, arXiv e-prints, arXiv:1902.01864

\bibitem[{{Mirizzi} {et~al.}(2016){Mirizzi}, {Tamborra}, {Janka}, {Saviano},
  {Scholberg}, {Bollig}, {H{\"u}depohl}, \&
  {Chakraborty}}]{2016NCimR..39....1M}
{Mirizzi}, A., {Tamborra}, I., {Janka}, H.-T., {et~al.} 2016, Nuovo Cimento
  Rivista Serie, 39, 1

\bibitem[{{Moe} \& {Di Stefano}(2017)}]{Moe:2017ApJS}
{Moe}, M., \& {Di Stefano}, R. 2017, \apjs, 230, 15

\bibitem[{{Moriya} \& {Langer}(2015)}]{2015A&A...573A..18M}
{Moriya}, T.~J., \& {Langer}, N. 2015, \aap, 573, A18

\bibitem[{{Moriya} {et~al.}(2018){Moriya}, {Tanaka}, {Yasuda}, {Jiang}, {Lee},
  {Maeda}, {Morokuma}, {Nomoto}, {Quimby}, {Suzuki}, {Takahashi}, {Tanaka},
  {Tominaga}, {Yamaguchi}, {Bernard}, {Cooke}, {Curtin}, {Galbany},
  {Gonzalez-Gaitan}, {Pignata}, {Pritchard}, {Komiyama}, \&
  {Lupton}}]{Moryia:2018arXiv180108240M}
{Moriya}, T.~J., {Tanaka}, M., {Yasuda}, N., {et~al.} 2018, arXiv e-prints,
  arXiv:1801.08240

\bibitem[{{Neijssel} {et~al.}(2019){Neijssel}, {Vigna-G{\'o}mez}, {Stevenson},
  {Barrett}, {Gaebel}, {Broekgaarden}, {de Mink}, {Sz{\'e}csi}, {Vinciguerra},
  \& {Mandel}}]{coen}
{Neijssel}, C.~J., {Vigna-G{\'o}mez}, A., {Stevenson}, S., {et~al.} 2019, arXiv
  e-prints, 1906.08136

\bibitem[{{Nicholl} {et~al.}(2013){Nicholl}, {Smartt}, {Jerkstrand}, {Inserra},
  {McCrum}, {Kotak}, {Fraser}, {Wright}, {Chen}, {Smith}, {Young}, {Sim},
  {Valenti}, {Howell}, {Bresolin}, {Kudritzki}, {Tonry}, {Huber}, {Rest},
  {Pastorello}, {Tomasella}, {Cappellaro}, {Benetti}, {Mattila}, {Kankare},
  {Kangas}, {Leloudas}, {Sollerman}, {Taddia}, {Berger}, {Chornock}, {Narayan},
  {Stubbs}, {Foley}, {Lunnan}, {Soderberg}, {Sanders}, {Milisavljevic},
  {Margutti}, {Kirshner}, {Elias-Rosa}, {Morales-Garoffolo}, {Taubenberger},
  {Botticella}, {Gezari}, {Urata}, {Rodney}, {Riess}, {Scolnic}, {Wood-Vasey},
  {Burgett}, {Chambers}, {Flewelling}, {Magnier}, {Kaiser}, {Metcalfe},
  {Morgan}, {Price}, {Sweeney}, \& {Waters}}]{Nicholl:2013Nature}
{Nicholl}, M., {Smartt}, S.~J., {Jerkstrand}, A., {et~al.} 2013, \nat, 502, 346

\bibitem[{{O'Leary} {et~al.}(2016){O'Leary}, {Meiron}, \&
  {Kocsis}}]{2016ApJ...824L..12O}
{O'Leary}, R.~M., {Meiron}, Y., \& {Kocsis}, B. 2016, \apjl, 824, L12

\bibitem[{{Paczynski}(1976)}]{1976IAUS...73...75P}
{Paczynski}, B. 1976, in IAU Symposium, Vol.~73, Structure and Evolution of
  Close Binary Systems, ed. P.~{Eggleton}, S.~{Mitton}, \& J.~{Whelan}, 75

\bibitem[{{Paczy{\'n}ski} \& {Sienkiewicz}(1972)}]{1972AcA....22...73P}
{Paczy{\'n}ski}, B., \& {Sienkiewicz}, R. 1972, Acta Astronomica, 22, 73

\bibitem[{{Pavlovskii} {et~al.}(2017){Pavlovskii}, {Ivanova}, {Belczynski}, \&
  {Van}}]{2017MNRAS.465.2092P}
{Pavlovskii}, K., {Ivanova}, N., {Belczynski}, K., \& {Van}, K.~X. 2017,
  \mnras, 465, 2092

\bibitem[{{Paxton} {et~al.}(2011){Paxton}, {Bildsten}, {Dotter}, {Herwig},
  {Lesaffre}, \& {Timmes}}]{Paxton:2011ApJS}
{Paxton}, B., {Bildsten}, L., {Dotter}, A., {et~al.} 2011, The Astrophysical
  Journal Supplement Series, 192, 3

\bibitem[{Paxton {et~al.}(2018)}]{Paxton:2017eie}
Paxton, B., {et~al.} 2018, Astrophys. J. Suppl., 234, 34

\bibitem[{{Peters}(1964)}]{1964PhRv..136.1224P}
{Peters}, P.~C. 1964, Physical Review, 136, 1224

\bibitem[{{Pols} {et~al.}(1998){Pols}, {Schr{\"o}der}, {Hurley}, {Tout}, \&
  {Eggleton}}]{Pols:1998MNRAS}
{Pols}, O.~R., {Schr{\"o}der}, K.-P., {Hurley}, J.~R., {Tout}, C.~A., \&
  {Eggleton}, P.~P. 1998, \mnras, 298, 525

\bibitem[{Powell {et~al.}(2019)Powell, Stevenson, Mandel, \&
  Tino}]{Powell:2019nmw}
Powell, J., Stevenson, S., Mandel, I., \& Tino, P. 2019, arXiv e-prints,
  arXiv:1905.04825

\bibitem[{Prajs {et~al.}(2017)}]{Prajs:2016cjj}
Prajs, S., {et~al.} 2017, Mon. Not. Roy. Astron. Soc., 464, 3568

\bibitem[{{Price-Whelan} {et~al.}(2018){Price-Whelan}, {Sip{\H{o}}cz},
  {G{\"u}nther}, {Lim}, {Crawford}, {Conseil}, {Shupe}, {Craig}, {Dencheva},
  {Ginsburg}, {VanderPlas}, {Bradley}, {P{\'e}rez-Su{\'a}rez}, {de Val-Borro},
  {Paper Contributors}, {Aldcroft}, {Cruz}, {Robitaille}, {Tollerud},
  {Coordination Committee}, {Ardelean}, {Babej}, {Bach}, {Bachetti}, {Bakanov},
  {Bamford}, {Barentsen}, {Barmby}, {Baumbach}, {Berry}, {Biscani}, {Boquien},
  {Bostroem}, {Bouma}, {Brammer}, {Bray}, {Breytenbach}, {Buddelmeijer},
  {Burke}, {Calderone}, {Cano Rodr{\'\i}guez}, {Cara}, {Cardoso}, {Cheedella},
  {Copin}, {Corrales}, {Crichton}, {D{\textquoteright}Avella}, {Deil},
  {Depagne}, {Dietrich}, {Donath}, {Droettboom}, {Earl}, {Erben}, {Fabbro},
  {Ferreira}, {Finethy}, {Fox}, {Garrison}, {Gibbons}, {Goldstein}, {Gommers},
  {Greco}, {Greenfield}, {Groener}, {Grollier}, {Hagen}, {Hirst}, {Homeier},
  {Horton}, {Hosseinzadeh}, {Hu}, {Hunkeler}, {Ivezi{\'c}}, {Jain}, {Jenness},
  {Kanarek}, {Kendrew}, {Kern}, {Kerzendorf}, {Khvalko}, {King}, {Kirkby},
  {Kulkarni}, {Kumar}, {Lee}, {Lenz}, {Littlefair}, {Ma}, {Macleod},
  {Mastropietro}, {McCully}, {Montagnac}, {Morris}, {Mueller}, {Mumford},
  {Muna}, {Murphy}, {Nelson}, {Nguyen}, {Ninan}, {N{\"o}the}, {Ogaz}, {Oh},
  {Parejko}, {Parley}, {Pascual}, {Patil}, {Patil}, {Plunkett}, {Prochaska},
  {Rastogi}, {Reddy Janga}, {Sabater}, {Sakurikar}, {Seifert}, {Sherbert},
  {Sherwood-Taylor}, {Shih}, {Sick}, {Silbiger}, {Singanamalla}, {Singer},
  {Sladen}, {Sooley}, {Sornarajah}, {Streicher}, {Teuben}, {Thomas},
  {Tremblay}, {Turner}, {Terr{\'o}n}, {van Kerkwijk}, {de la Vega}, {Watkins},
  {Weaver}, {Whitmore}, {Woillez}, {Zabalza}, \& {Contributors}}]{astropy:2018}
{Price-Whelan}, A.~M., {Sip{\H{o}}cz}, B.~M., {G{\"u}nther}, H.~M., {et~al.}
  2018, \aj, 156, 123

\bibitem[{Quimby {et~al.}(2013)Quimby, Yuan, Akerlof, \&
  Wheeler}]{Quimby:2013jb}
Quimby, R.~M., Yuan, F., Akerlof, C., \& Wheeler, J.~C. 2013, Mon. Not. Roy.
  Astron. Soc., 431, 912

\bibitem[{{Quimby} {et~al.}(2011){Quimby}, {Kulkarni}, {Kasliwal}, {Gal- Yam},
  {Arcavi}, {Sullivan}, {Nugent}, {Thomas}, {Howell}, {Nakar}, {Bildsten},
  {Theissen}, {et~al.}}]{Quimby:2011Nature}
{Quimby}, R.~M., {Kulkarni}, S.~R., {Kasliwal}, M.~M., {et~al.} 2011, \nat,
  474, 487

\bibitem[{{Raghavan} {et~al.}(2010){Raghavan}, {McAlister}, {Henry}, {Latham},
  {Marcy}, {Mason}, {Gies}, {White}, \& {ten Brummelaar}}]{Raghavan:2010ApJS}
{Raghavan}, D., {McAlister}, H.~A., {Henry}, T.~J., {et~al.} 2010, \apjs, 190,
  1

\bibitem[{{Rastello} {et~al.}(2019){Rastello}, {Amaro-Seoane}, {Arca-Sedda},
  {Capuzzo- Dolcetta}, {Fragione}, \& {Tosta e Melo}}]{Rastello:2019}
{Rastello}, S., {Amaro-Seoane}, P., {Arca-Sedda}, M., {et~al.} 2019, \mnras,
  483, 1233

\bibitem[{{Renzo} {et~al.}(2017){Renzo}, {Ott}, {Shore}, \& {de
  Mink}}]{Renzo:2017}
{Renzo}, M., {Ott}, C.~D., {Shore}, S.~N., \& {de Mink}, S.~E. 2017, \aap, 603,
  A118

\bibitem[{{Repetto} {et~al.}(2012){Repetto}, {Davies}, \&
  {Sigurdsson}}]{2012MNRAS.425.2799R}
{Repetto}, S., {Davies}, M.~B., \& {Sigurdsson}, S. 2012, \mnras, 425, 2799

\bibitem[{Repetto {et~al.}(2017)Repetto, Igoshev, \&
  Nelemans}]{Repetto:2017gry}
Repetto, S., Igoshev, A.~P., \& Nelemans, G. 2017, Mon. Not. Roy. Astron. Soc.,
  467, 298

\bibitem[{{Rodriguez} {et~al.}(2018){Rodriguez}, {Amaro-Seoane}, {Chatterjee},
  \& {Rasio}}]{2018PhRvL.120o1101R}
{Rodriguez}, C.~L., {Amaro-Seoane}, P., {Chatterjee}, S., \& {Rasio}, F.~A.
  2018, Physical Review Letters, 120, 151101

\bibitem[{Rodriguez \& Antonini(2018)}]{Rodriguez:2018jqu}
Rodriguez, C.~L., \& Antonini, F. 2018, Astrophys. J., 863, 7

\bibitem[{Rodriguez {et~al.}(2016)Rodriguez, Haster, Chatterjee, Kalogera, \&
  Rasio}]{Rodriguez:2016avt}
Rodriguez, C.~L., Haster, C.-J., Chatterjee, S., Kalogera, V., \& Rasio, F.~A.
  2016, Astrophys. J., 824, L8

\bibitem[{{Roulet} \& {Zaldarriaga}(2019)}]{Roulet:2018jbe}
{Roulet}, J., \& {Zaldarriaga}, M. 2019, \mnras, 484, 4216

\bibitem[{Samsing \& Ilan(2019)}]{Samsing:2017xod}
Samsing, J., \& Ilan, T. 2019, Mon. Not. Roy. Astron. Soc., 482, 30

\bibitem[{{Sana} {et~al.}(2012){Sana}, {de Mink}, {de Koter}, {Langer},
  {Evans}, {Gieles}, {Gosset}, {Izzard}, {Le Bouquin}, \&
  {Schneider}}]{Sana:2012Sci}
{Sana}, H., {de Mink}, S.~E., {de Koter}, A., {et~al.} 2012, Science, 337, 444

\bibitem[{{Sanyal} {et~al.}(2015){Sanyal}, {Grassitelli}, {Langer}, \&
  {Bestenlehner}}]{Sanyal:2015A&A}
{Sanyal}, D., {Grassitelli}, L., {Langer}, N., \& {Bestenlehner}, J.~M. 2015,
  \aap, 580, A20

\bibitem[{{Sanyal} {et~al.}(2017){Sanyal}, {Langer}, {Sz{\'e}csi}, {-C Yoon},
  \& {Grassitelli}}]{Sanyal:2017A&A}
{Sanyal}, D., {Langer}, N., {Sz{\'e}csi}, D., {-C Yoon}, S., \& {Grassitelli},
  L. 2017, \aap, 597, A71

\bibitem[{{Schneider} {et~al.}(2018){Schneider}, {Sana}, {Evans},
  {Bestenlehner}, {Castro}, {Fossati}, {Gr{\"a}fener}, {Langer},
  {Ram{\'\i}rez-Agudelo}, {Sab{\'\i}n-Sanjuli{\'a}n}, {Sim{\'o}n-D{\'\i}az},
  {Tramper}, {Crowther}, {de Koter}, {de Mink}, {Dufton}, {Garcia}, {Gieles},
  {H{\'e}nault- Brunet}, {Herrero}, {Izzard}, {Kalari}, {Lennon}, {Ma{\'\i}z
  Apell{\'a}niz}, {Markova}, {Najarro}, {Podsiadlowski}, {Puls}, {Taylor}, {van
  Loon}, {Vink}, \& {Norman}}]{2018Sci...359...69S}
{Schneider}, F.~R.~N., {Sana}, H., {Evans}, C.~J., {et~al.} 2018, Science, 359,
  69

\bibitem[{Sigurdsson \& Hernquist(1993)}]{Sigurdsson:1993zrm}
Sigurdsson, S., \& Hernquist, L. 1993, Nature, 364, 423

\bibitem[{Spera \& Mapelli(2017)}]{Spera:2017fyx}
Spera, M., \& Mapelli, M. 2017, Mon. Not. Roy. Astron. Soc., 470, 4739

\bibitem[{{Spera} {et~al.}(2015){Spera}, {Mapelli}, \&
  {Bressan}}]{Spera:2015MNRAS}
{Spera}, M., {Mapelli}, M., \& {Bressan}, A. 2015, \mnras, 451, 4086

\bibitem[{{Spera} {et~al.}(2019){Spera}, {Mapelli}, {Giacobbo}, {Trani},
  {Bressan}, \& {Costa}}]{Spera:2018wnw}
{Spera}, M., {Mapelli}, M., {Giacobbo}, N., {et~al.} 2019, \mnras, 485, 889

\bibitem[{Stevenson {et~al.}(2015)Stevenson, Ohme, \&
  Fairhurst}]{Stevenson:2015bqa}
Stevenson, S., Ohme, F., \& Fairhurst, S. 2015, Astrophys. J., 810, 58

\bibitem[{{Stevenson} {et~al.}(2017){Stevenson}, {Vigna-G{\'o}mez}, {Mandel},
  {Barrett}, {Neijssel}, {Perkins}, \& {de
  Mink}}]{Stevenson2017FormationEvolution}
{Stevenson}, S., {Vigna-G{\'o}mez}, A., {Mandel}, I., {et~al.} 2017, Nature
  Communications, 8, 14906

\bibitem[{Stone {et~al.}(2017)Stone, Metzger, \& Haiman}]{Stone:2016wzz}
Stone, N.~C., Metzger, B.~D., \& Haiman, Z. 2017, Mon. Not. Roy. Astron. Soc.,
  464, 946

\bibitem[{{Sz{\'e}csi} {et~al.}(2015){Sz{\'e}csi}, {Langer}, {Yoon}, {Sanyal},
  {de Mink}, {Evans}, \& {Dermine}}]{Szecsi:2015A&A}
{Sz{\'e}csi}, D., {Langer}, N., {Yoon}, S.-C., {et~al.} 2015, \aap, 581, A15

\bibitem[{Talbot \& Thrane(2018)}]{Talbot:2018cva}
Talbot, C., \& Thrane, E. 2018, Astrophys. J., 856, 173

\bibitem[{{The LIGO Scientific Collaboration} {et~al.}(2018{\natexlab{a}}){The
  LIGO Scientific Collaboration}, {the Virgo Collaboration}, {Abbott},
  {Abbott}, {Abbott}, {Abraham}, {Acernese}, {Ackley}, {Adams}, {Adhikari},
  {et~al.}}]{Collaboration2018BinaryVirgoa}
{The LIGO Scientific Collaboration}, {the Virgo Collaboration}, {Abbott},
  B.~P., {et~al.} 2018{\natexlab{a}}, arXiv e-prints, arXiv:1811.12940

\bibitem[{{The LIGO Scientific Collaboration} {et~al.}(2018{\natexlab{b}}){The
  LIGO Scientific Collaboration}, {the Virgo Collaboration}, {Abbott},
  {Abbott}, {Abbott}, {Abraham}, {Acernese}, {Ackley}, {Adams}, {Adhikari},
  {Adya}, {et~al.}}]{Collaboration2018GWTC-1:Runsa}
---. 2018{\natexlab{b}}, arXiv e-prints, arXiv:1811.12907

\bibitem[{{Tout} {et~al.}(1996){Tout}, {Pols}, {Eggleton}, \&
  {Han}}]{1996MNRAS.281..257T}
{Tout}, C.~A., {Pols}, O.~R., {Eggleton}, P.~P., \& {Han}, Z. 1996, \mnras,
  281, 257

\bibitem[{van~den Heuvel {et~al.}(2017)van~den Heuvel, Portegies~Zwart, \&
  de~Mink}]{Heuvel:2017sfe}
van~den Heuvel, E. P.~J., Portegies~Zwart, S.~F., \& de~Mink, S.~E. 2017, Mon.
  Not. Roy. Astron. Soc., 471, 4256

\bibitem[{Venumadhav {et~al.}(2019)Venumadhav, Zackay, Roulet, Dai, \&
  Zaldarriaga}]{Venumadhav:2019lyq}
Venumadhav, T., Zackay, B., Roulet, J., Dai, L., \& Zaldarriaga, M. 2019, arXiv
  e-prints, arXiv:1904.07214

\bibitem[{{Vigna-G{\'o}mez} {et~al.}(2019){Vigna-G{\'o}mez}, {Justham},
  {Mandel}, {de Mink}, \& {Podsiadlowski}}]{Vigna-Gomez:2019sky}
{Vigna-G{\'o}mez}, A., {Justham}, S., {Mandel}, I., {de Mink}, S.~E., \&
  {Podsiadlowski}, P. 2019, \apjl, 876, L29

\bibitem[{{Vigna-G{\'o}mez} {et~al.}(2018){Vigna-G{\'o}mez}, {Neijssel},
  {Stevenson}, {Barrett}, {Belczynski}, {Justham}, {de Mink}, {M{\"u}ller},
  {Podsiadlowski}, {Renzo}, {Sz{\'e}csi}, \& {Mandel}}]{2018MNRAS.481.4009V}
{Vigna-G{\'o}mez}, A., {Neijssel}, C.~J., {Stevenson}, S., {et~al.} 2018,
  \mnras, 481, 4009

\bibitem[{Vink {et~al.}(2001)Vink, de~Koter, \& Lamers}]{Vink:2001cg}
Vink, J.~S., de~Koter, A., \& Lamers, H. J. G. L.~M. 2001, Astron. Astrophys.,
  369, 574

\bibitem[{{Webbink}(1984)}]{1984ApJ...277..355W}
{Webbink}, R.~F. 1984, \apj, 277, 355

\bibitem[{Woosley(2017)}]{Woosley:2016hmi}
Woosley, S.~E. 2017, Astrophys. J., 836, 244

\bibitem[{Woosley(2018)}]{Woosley:2018wuj}
---. 2018, Astrophys. J., 863, 105

\bibitem[{{Woosley}(2019)}]{Woosley2019TheLoss}
{Woosley}, S.~E. 2019, arXiv e-prints, arXiv:1901.00215

\bibitem[{Woosley {et~al.}(2007)Woosley, Blinnikov, \& Heger}]{Woosley:2007qp}
Woosley, S.~E., Blinnikov, S., \& Heger, A. 2007, Nature, 450, 390

\bibitem[{{Woosley} {et~al.}(2002){Woosley}, {Heger}, \&
  {Weaver}}]{Woosley2002TheStars}
{Woosley}, S.~E., {Heger}, A., \& {Weaver}, T.~A. 2002, Reviews of Modern
  Physics, 74, 1015

\bibitem[{Wysocki {et~al.}(2018{\natexlab{a}})Wysocki, Gerosa, O'Shaughnessy,
  Belczynski, Gladysz, Berti, Kesden, \& Holz}]{Wysocki:2017isg}
Wysocki, D., Gerosa, D., O'Shaughnessy, R., {et~al.} 2018{\natexlab{a}}, Phys.
  Rev., D97, 043014

\bibitem[{Wysocki {et~al.}(2018{\natexlab{b}})Wysocki, Lange, \&
  O.~'shaughnessy}]{Wysocki:2018mpo}
Wysocki, D., Lange, J., \& O.~'shaughnessy, R. 2018{\natexlab{b}}, arXiv
  e-prints, arXiv:1805.06442

\bibitem[{{Xu} \& {Li}(2010)}]{2010ApJ...716..114X}
{Xu}, X.-J., \& {Li}, X.-D. 2010, \apj, 716, 114

\bibitem[{{Yoon} {et~al.}(2012){Yoon}, {Dierks}, \& {Langer}}]{Yoon:2012A&A}
{Yoon}, S.-C., {Dierks}, A., \& {Langer}, N. 2012, \aap, 542, A113

\bibitem[{{Yoshida} {et~al.}(2016){Yoshida}, {Umeda}, {Maeda}, \&
  {Ishii}}]{Yoshida:2016MNRAS}
{Yoshida}, T., {Umeda}, H., {Maeda}, K., \& {Ishii}, T. 2016, \mnras, 457, 351

\bibitem[{Zackay {et~al.}(2019)Zackay, Venumadhav, Dai, Roulet, \&
  Zaldarriaga}]{Zackay:2019tzo}
Zackay, B., Venumadhav, T., Dai, L., Roulet, J., \& Zaldarriaga, M. 2019, arXiv
  e-prints, arXiv:1902.10331

\bibitem[{Zapartas {et~al.}(2017)}]{Zapartas:2017zsb}
Zapartas, E., {et~al.} 2017, Astron. Astrophys., 601, A29

\bibitem[{Zevin {et~al.}(2017)Zevin, Pankow, Rodriguez, Sampson, Chase,
  Kalogera, \& Rasio}]{Zevin:2017evb}
Zevin, M., Pankow, C., Rodriguez, C.~L., {et~al.} 2017, Astrophys. J., 846, 82

\end{thebibliography}


\end{document}